# A Graphical Theoretical Framework for Cylindrical Cavity Expansion in Mohr-Coulomb Geomaterials

**SHENG-LI CHEN**

Associate Professor
Department of Civil & Environmental Engineering
Louisiana State University
Baton Rouge, LA, USA
Tel: +1 (225) 578 2432
Email: shenglichen@lsu.edu

**Abstract:** This paper develops a complete analytical solution for the drained (or dry) cylindrical cavity expansion in non-associated Mohr-Coulomb soil, by using the graphical approach and Lagrangian formulation of the cavity boundary value problem (through tracing the responses of a single soil particle at the cavity wall). The novelty of the new solution lies not only in the relaxation of the strict intermediacy assumption for the vertical stress as usually adopted in the previous analyses, but in the comprehensive consideration of arbitrary values of $K_0$, the coefficient of earth pressure at rest, as well. The essence of the so-called graphical method, i.e., the unique geometrical analysis and tracking of the deviatoric stress trajectory, is fulfilled by leveraging the deformation requirement that during drained expansion the progressive development of the radial and tangential strains must maintain to be compressive and tensile, respectively. With the incorporation of the radial equilibrium condition, the problem is formulated to solve a single first-order differential equation for the internal cavity pressure with respect to a pivotal auxiliary variable, for all the distinct scenarios of $K_0$ being covered. Some selected results are presented for the calculated cavity expansion curve and limit cavity pressure through an example analysis.

**Introduction**

Analysis of cylindrical cavity expansion in a soil mass provides a versatile and reasonably accurate geomechanics approach for the study of many important problems in geotechnical engineering [1], notably the interpretation of pressuremeter/cone penetration tests for determination of key soil properties in situ and the prediction of driven pile capacity. In the last 40 years, tremendous research efforts have been conducted on the cavity expansion models, with significant progress being made in the development of analytical/semi-analytical solutions for such a fundamental boundary value problem [2-12].

One type of elastoplastic constitutive models most commonly adopted in the derivation of cavity expansion solutions is the Mohr-Coulomb model suitable for cohesive-frictional soils [2, 3, 13-18]. However, the major limitations of the numerous existing Mohr-Coulomb solutions pertaining to the cylindrical cavity expansion are that the axial stress has been almost exclusively formulated as an intermediate one with respect to the other two of radial and tangential stresses, and that the soil mass is assumed to be under a hydrostatic in situ stress condition. The drawback of the former assumption, as has long been recognized [19], is obvious, since the intermediacy of the axial stress does not necessarily remain valid throughout the cavity expansion process but instead depending on the values of Poisson's ratio and the angle of internal friction [3, 20-22]. This is particularly true when the general case is encountered in geotechnical practice that the in situ stresses are non-isotropic with $K_0$, the coefficient of earth pressure at rest, not equal to unity.

There exist only few exceptions to addressing the above-mentioned limitations, in primarily the analysis of cavity excavation/contraction problem in rock mechanics, as reported by Florence & Schwer [20] and Reed [22], where the involvement of the axial stress in the Mohr-Coulomb criterion, and consequently the plastic strain component in the vertical direction, are accounted for to some extent. However, the former of these two contributions has incorporated only one of the possible cases in regards to the yield surface and to the initial yielding conditions involved, and used an associated flow rule; while the latter was limited to a simple scenario with the rock mass being subjected to a hydrostatic in situ stress field, i.e., the earth pressure coefficient $K_0 = 1$. Moreover, both the two analyses adopted the small-strain theory, which might not be capable of capturing the large deformation effect of the plastic state/zone in the modelling of cavity expansion. Another noteworthy work taking into consideration the influence of out-of-plane stress on the cavity/hole responses was due to Risnes et al. [21], who studied theoretically the steady-state stress distributions in a saturated, poorly consolidated sand around a drilled wellbore, following basically the same solution procedure as outlined in Florence & Schwer [20]. The solution of Risnes et al. [21] nevertheless was still based on the assumption of small-strain deformation with an associated flow rule. And it is believed that the two possible stress ordering scenarios at the elastic/plastic boundary covered in their formulations are incomplete as the radial stress component at the onset of yielding may become an intermediate one and needs not always to be the smallest.

In an effort to derive a rigorous and complete solution for the undrained cavity expansion in Mohr-Coulomb soil, the author (Chen & Wang [23]) most recently proposed a new theoretical framework, based on the so-called graphical analysis approach [8] and on the Lagrangian description [7] via tracking merely the responses of a soil particle at the cavity wall throughout the analysis. In the formulations of Chen & Wang [23], the mathematical difficulties involved with the flow rate calculation corresponding to a stress state lying on the edge of two adjacent (Mohr-Coulomb) yield surfaces have been effectively treated by using the generalized Koiter's [24] theory for non-associated plasticity [25]. In particular, under the undrained conditions with vanishing volumetric strain, the graphical procedure allows the effective stress path pertaining to the cylindrical expansion problem to be diagrammatically determined in the most convenient and efficient way. It thus makes possible the definitive identification of the flow rule/stiffness matrix, which evolve continuously with the position of the current stress state (relative to the multiple yield surfaces of the Mohr-Coulomb failure criterion) as loading proceeds.

For relatively permeable sandy or silty soils, significant amount of drainage may occur during the cavity expansion. Hence the soil deformation is generally not equivoluminal, and in such a case the cavity analysis may be reasonably performed under drained conditions. However, in the drained situation, the strong condition of deviatoric strain increment being simply horizontal resulting from the zero volumetric strain constraint (as demanded in the undrained case) is no longer valid, which, therefore, is expected to impose great difficulties in determining the stress path. This paper attempts to establish a graphical analysis framework for the challenging problem of drained cavity expansion in Mohr-Coulomb soil, with the comprehensive consideration of the influences of the vertical stress for arbitrary values of $K_0$. It differs substantially from the previous undrained solution [23] in three main respects. Firstly, the essential stress path in the deviatoric plane has to be geometrically determined in a distinct and much subtle/innovative manner, due to the requirement of volume change involved in the drained analysis. Secondly, the developed drained cavity expansion solution indeed provides full coverage of $K_0$ values, making it a first complete and unique one of its kind in the literature. Lastly, the (total) radial stress, or cavity pressure, needs to be implicitly evaluated through numerically solving a first-order differential equation (derived from the constitutive relationship coupled with the Lagrangian form of the equilibrium equation), rather than being explicitly expressible as in the undrained circumstance.

It is interesting to note and worth mentioning that, even with the relaxation of the zero volumetric strain condition, the stress paths for the present drained case can still be effectively tracked by leveraging the deformation characteristic of the cylindrical expansion problem that the corresponding radial, tangential, and vertical strain increments must remain compressive, tensile, and vanishing, respectively. Nevertheless, for relatively large or small values of $K_0$ (with respect to unity), the stress paths cannot be directly determined graphically without being supplied by the equation of radial equilibrium, and they are found to be actually curved lines in the deviatoric plane if the vertical stress becomes no longer the intermediate principal stress. Some selected numerical results are presented to cover different values of $K_0$ equal to, less or greater than 1, and to demonstrate the influences of the friction angle on the cavity expansion curve and the calculated limit pressure as well.

## Graphical method for drained cavity expansion in Mohr-Coulomb soil

In this section the graphical procedure proposed by Chen & Wang [23] for the undrained cylindrical cavity expansion in non-associated Mohr-Coulomb soil will be significantly expanded to solve the more difficult cavity boundary value problem under the drained conditions. To provide a general and complete solution to the problem, the following three case scenarios concerning the magnitude of the earth pressure coefficient, i.e., $K_0 = 1$, $K_0 < 1$, and $K_0 > 1$, will be fully covered in the development of the graphical analysis approach.

### HYDROSTATIC IN-SITU STRESS CONDITION WITH $K_0 = 1$

Consider first the specific expansion case of a cylindrical cavity in a Mohr-Coulomb material under hydrostatic in situ stress field, i.e., $K_0$ equal to unity. Let $a_0$ be the initial radius of the cavity, which is expanded to its current radius $a$ when the internal cavity pressure is increased to $\sigma_a$, and let $\sigma_h$ and $\sigma_v$ be the in situ horizontal and vertical stresses, respectively. For the case of $K_0 = 1$, $\sigma_h = \sigma_v = p_0$, where $p_0$ denotes the mean stress at rest acting throughout the soil.

Note that in Chen & Wang [23] for the undrained cavity analysis with an initial isotropic stress state ($K_0 = 1$), the key component of the graphical approach developed is to represent and/or decompose the stress and strain states/increments of a material point in both the deviatoric stress and strain planes. For the case of drained expansion, the volumetric strain nevertheless is not equal to zero, so the vector of deviatoric strain increment $D\boldsymbol{e}$ must not be oriented in the horizontal direction (with Lode angle $\theta = \frac{5\pi}{3}$) in the deviatoric strain plane. This indicates that the formulations and analysis adopted in Chen & Wang [23] for the graphical determination of the stress paths will fall short when applied to the present drained case. Fortunately, for the Mohr-Coulomb plasticity model specified by multiple planar yield surfaces, it will be found possible for the deviatoric stress path to be still graphically deduced via a distinctively different and somewhat subtle procedure (note: only the geometrical consideration of the stress state is virtually needed if the stress path is positioned in the main sextant of $\frac{3\pi}{2} < \theta < \frac{11\pi}{6}$). This essentially constitutes one of the major contributions of this work, as described below.

**(a) Case I: $2\nu \geq 1 - \sin\phi$**

Consider now the case that Poisson's ratio $\nu$ and the friction angle $\phi$ of the soil satisfy the condition $2\nu \geq 1 - \sin\phi$. As similar to Chen & Wang [23], let $\mathbf{X_0X_{ep}XX_u}$ represents the typical deviatoric stress path of a soil particle at the cavity surface, see Fig. 1(a). Here the point $\mathbf{X}(s_r, s_\theta, s_z)$ represents the current stress state projected on the $\pi$-plane, where $s_r$, $s_\theta$, and $s_z$ denote the stress deviator components in radial, tangential, and vertical principal directions, respectively, with the sign convention of compression being positive; $\mathbf{X_0}$ coinciding with the origin O is the initial deviatoric stress point; while $\mathbf{X_{ep}}$ and $\mathbf{X_u}$ correspond respectively to the $\pi$-plane projections of the elastic-plastic transition stress state and the ultimate stress state.

For $K_0 = 1$, the cavity response during the elastic phase of a cylindrical expansion under drained conditions requires that $\sigma_z = \frac{1}{2}(\sigma_r + \sigma_\theta)$ [7], where $\sigma_r$, $\sigma_\theta$, and $\sigma_z$ are the radial, tangential, and vertical stress components. Therefore, the segment of the elastic stress path $\mathbf{X_0 X_{ep}}$ must be a horizontal straight line in alignment with the $x$ axis [8], as plotted in Fig. 1(a). Once the stress path hits the Mohr-Coulomb yield surface (fixed in the principal stress space for perfectly plastic model) at point $\mathbf{X_{ep}}$, the soil particle at the cavity wall will start to undergo plastic deformations. Since $\mathbf{X_{ep}}$ lies in the sector of $\frac{3\pi}{2} < \theta < \frac{11\pi}{6}$ corresponding to the ordering of the principal stresses, $\sigma_r > \sigma_z > \sigma_\theta$, an infinitesimal amount of stress vector increment over this very stress state, say $D\boldsymbol{\sigma}_{ep} = \{D\sigma_{r,ep}, D\sigma_{\theta,ep}, D\sigma_{z,ep}\}^T$ (not shown but whose projection on the deviatoric plane is represented as $D\boldsymbol{s}_{ep}$ in the figure), needs to be located on the same face of the yield surfaces

$$F_{r\theta}(\sigma_r, \sigma_\theta) = \frac{\sigma_r - \sigma_\theta}{2} - \frac{\sigma_r + \sigma_\theta}{2}\sin\phi - c\cos\phi \qquad (1)$$

and satisfy the consistency condition $DF_{r\theta} = 0$, i.e.,

$$D\sigma_{\theta,ep} = \frac{1}{\alpha} D\sigma_{r,ep} \qquad (2)$$

where $F_{r\theta}$ denotes the yield function that depends only on the major principal stress $\sigma_r$ and the minor principal stress $\sigma_\theta$; $c$ is the cohesion of the soil; and $\alpha = \frac{1+\sin\phi}{1-\sin\phi}$, as similar with Yu & Houlsby [3], is a constant introduced to abbreviate the mathematical formulation.

Eq. (2) establishes a simple, linear relationship between the two incremental stress components $D\sigma_{\theta,ep}$ and $D\sigma_{r,ep}$. In fact, they are also linearly related to the third stress increment $D\sigma_{z,ep}$ through the restriction imposed by the plane strain condition of $D\varepsilon_z = 0$, as is shown next. Here $D\varepsilon_z$ denotes the vertical total strain increment and can be split into elastic, $D\varepsilon_z^e$, and plastic, $D\varepsilon_z^p$, parts to give

$$D\varepsilon_z = D\varepsilon_z^e + D\varepsilon_z^p = 0 \tag{3}$$

in which

$$D\varepsilon_z^e = \frac{1}{2G(1+v)}(D\sigma_{z,ep} - vD\sigma_{r,ep} - vD\sigma_{\theta,ep}) \tag{4}$$

according to Hooke's law, with $G$ being the shear modulus; while

$$D\varepsilon_z^p = \lambda_{r\theta} \frac{\partial P_{r\theta}(\sigma_r, \sigma_\theta)}{\partial \sigma_z} \tag{5}$$

where $P_{r\theta}$ denotes the plastic potential function passing through the current transition stress state $\mathbf{X_{ep}}$ (see the dashed hexagon in Fig. 1(a)). For a dilatant non-associated Mohr-Coulomb model, it takes a similar form to that of the yield function ($F_{r\theta}$) as follows [26]

$$P_{r\theta}(\sigma_r, \sigma_\theta) = \frac{\sigma_r - \sigma_\theta}{2} - \frac{\sigma_r + \sigma_\theta}{2}\sin\psi - c_{r\theta}^* \cos\psi \tag{6}$$

where $\psi$ is known as the angle of dilation and $c_{r\theta}^*$ is a constant parameter reliant on the point $\mathbf{X_{ep}}$; and $\lambda_{r\theta}$ is a scalar factor of proportionality corresponding to the potential function $P_{r\theta}$.

Since $P_{r\theta}$ is not expressed as a function of the intermediate stress $\sigma_z$ explicitly, it follows from Eq. (5) that $D\varepsilon_z^p = 0$. This result substituted in Eq. (3) and then Eq. (4) yields

$$D\sigma_{z,ep} = v(D\sigma_{r,ep} + D\sigma_{\theta,ep}) \tag{7}$$

which in combination with Eq. (2) leads to

$$\begin{Bmatrix} D\sigma_{r,ep} \\ D\sigma_{\theta,ep} \\ D\sigma_{z,ep} \end{Bmatrix} = \begin{Bmatrix} 1 \\ \frac{1}{\alpha} \\ v(1 + \frac{1}{\alpha}) \end{Bmatrix} D\sigma_{r,ep} \tag{8}$$

This equation identifies the orientation of the stress increment vector $D\boldsymbol{\sigma}_{ep}$ at point $\mathbf{X_{ep}}$ and thus its projection $D\boldsymbol{s}_{ep}$ on the $\pi$-plane, from which the Lode angle of $D\boldsymbol{s}_{ep}$ can be determined as (see Fig. 1(a))

$$\theta_{D\boldsymbol{s}_{ep}} = 2\pi + \tan^{-1}\left(\frac{1}{\sqrt{3}} \frac{1 - 3\sin\phi - 2v}{1 - 2v + \sin\phi}\right) \tag{9}$$

Eq. (8) can also be rearranged to give

$$D\sigma_{z,ep} = \nu(1+\frac{1}{\alpha})D\sigma_{r,ep} \leq \frac{1}{2}(1+\frac{1}{\alpha})D\sigma_{r,ep} = \frac{1}{2}(D\sigma_{r,ep} + D\sigma_{\theta,ep}) \tag{10a}$$

$$Dp_{ep} = \frac{1}{3}(D\sigma_{r,ep} + D\sigma_{\theta,ep} + D\sigma_{z,ep}) = \frac{1}{3}(1+\nu)(1+\frac{1}{\alpha})D\sigma_{r,ep} \geq 0 \tag{10b}$$

where $Dp_{ep}$ denotes the mean stress increment at point $\mathbf{X_{ep}}$ and where use has been made of the obvious condition $D\sigma_{r,ep} \geq 0$. The above two equations indicate that $D\mathbf{s}_{ep}$ must be constrained to lie between the two lines of $\mathbf{X_{ep}}\mathrm{x}$ ($\theta = \frac{5\pi}{3}$) and $\mathbf{X_{ep}}\mathrm{B}$ ($\theta_{\mathbf{X_{ep}B}} = 2\pi + \tan^{-1}\left[\frac{\sqrt{3}(1-\sin\phi)}{3+\sin\phi}\right]$), i.e., the shaded region as shown in Fig. 1(a).

From Eq. (9), it is found that the direction of the deviatoric stress path $D\mathbf{s}_{ep}$, i.e., the Lode angle $\theta_{D\mathbf{s}_{ep}}$, is independent of the current stress state of the soil and controlled only by the two constant parameters of $\nu$ and $\phi$. Such a finding applies with the continued cavity loading and therefore with the arbitrary stress point $\mathbf{X}$ and its corresponding stress increment $D\mathbf{s}$, as long as $\sigma_r > \sigma_\theta > \sigma_z$ is satisfied in the plastic deformation (Fig. 1(a)). Hence, $\theta_{D\mathbf{s}} = \theta_{D\mathbf{s}_{ep}}$ and $\mathbf{X_{ep}X}$ turns out to be a straight line. Note that this desirable conclusion as well as Eqs. (8) and (9) for the expressions of stress increment vector and Lode angle are exactly the same as those for the undrained cavity expansion [23], although they nevertheless have to be deduced based on a quite different procedure with no involvement of/reliance on the geometrical representation of the deviatoric strain increment. For the present case of $2\nu \geq 1 - \sin\phi$, one can easily prove that the value of $\theta_{D\mathbf{s}_{ep}}$ (or $\theta_{D\mathbf{s}}$) calculated from Eq. (9) will always fall into a narrower range of $\frac{5\pi}{3} \leq \theta \leq \frac{11\pi}{6}$, the latter equal sign corresponding to the extreme case of $2\nu = 1 - \sin\phi$. This clearly means that the plastic stress path $\mathbf{X_{ep}XX_u}$, as with the undrained case, will never meet the $s_r$ axis in the $\pi$-plane so that $\sigma_z$ indeed remains to be the intermediate stress during the entire cavity expansion process. The analysis of stress/deformation responses of the cavity therefore should be carried out by using Eqs. (1) and (6) given above as single yield and potential functions, which will now be described.

Following Chen & Wang [23], when the current stress state **X** is situated on the specific Mohr-Coulomb yield surface defined by Eq. (1), the elastoplastic constitutive equation can be written as (there is no need to distinguish between effective and total stresses for drained analysis)

$$\begin{Bmatrix} D\sigma_r \\ D\sigma_\theta \\ D\sigma_z \end{Bmatrix} = \frac{1}{\Delta} \begin{bmatrix} b_{11} & b_{12} & b_{13} \\ b_{21} & b_{22} & b_{23} \\ b_{31} & b_{32} & b_{33} \end{bmatrix} \begin{Bmatrix} D\varepsilon_r \\ D\varepsilon_\theta \\ D\varepsilon_z \end{Bmatrix} \quad (11)$$

where $D\sigma_r$, $D\sigma_\theta$, and $D\sigma_z$ are the stress increments in the radial, tangential, and vertical directions, respectively; $D\varepsilon_r$, $D\varepsilon_\theta$, and the already defined $D\varepsilon_z$ are the corresponding strain increments; $\Delta = \frac{1-2\nu}{G}$; and $b_{ij}(i,j = 1, 2, 3)$ are all constants depending explicitly on $G$, $\nu$, $\phi$, and $\psi$, which have the same definitions as those for the undrained case [23].

Eq. (11) seemingly has provided three incremental stress-strain relations. However, only one of them are independent, attributing to the fact that the three stress increments, $D\sigma_r$, $D\sigma_\theta$, and $D\sigma_z$, have to meet the requirements imposed by Eqs. (2) and (7). Here the first row of Eq. (11) will be used, which, by eliminating $D\varepsilon_z$, becomes

$$D\sigma_r = \frac{1}{\Delta}[b_{11}D\varepsilon_v + (b_{12} - b_{11})D\varepsilon_\theta] \quad (12)$$

where $D\varepsilon_v = D\varepsilon_r + D\varepsilon_\theta$ is the increment of the volumetric strain; and the incremental tangential strain $D\varepsilon_\theta$, on account of the effect of large deformations, takes the form of $D\varepsilon_\theta = -\frac{Dr}{r}$, with $Dr$ denoting the infinitesimal change in radial position of the soil particle.

The above constitutive relation (12) shall be solved together with the radial equilibrium condition

$$\frac{\partial \sigma_r}{\partial r} + \frac{\sigma_r - \sigma_\theta}{r} = 0 \tag{13}$$

to finally obtain a complete solution for the cavity responses. A convenient way to achieve this is to convert the Eulerian-based governing equation (13) to a Lagrangian form, which can be accomplished by employing the auxiliary variable method originally proposed by Chen [27] and Chen & Abousleiman [7]. With the introduction of a pivotal auxiliary variable $\xi$ such that $\xi = \frac{a-a_0}{a}$, the two expressions (12) and (13) transform into

$$\frac{D\sigma_a}{D\xi} = -\frac{\sigma_a - \sigma_{\theta,a}}{1 - \xi - \frac{e^{\varepsilon_{v,a}}}{1-\xi}} \tag{14}$$

$$\frac{D\varepsilon_{v,a}}{D\xi} = -\frac{\Delta}{b_{11}}\left\{-\frac{D\sigma_a}{D\xi} + \frac{b_{11}-b_{12}}{\Delta(1-\xi)}\right\} \tag{15}$$

where $\sigma_a$, $\sigma_{\theta,a}$, and $\varepsilon_{v,a}$ represent the radial and tangential stresses, and the volumetric strain at the cavity wall, respectively.

On performing the integration of Eq. (15), and substituting the resulting $\varepsilon_{v,a}$ in Eq. (14), gives

$$\frac{D\sigma_a}{D\xi} = -\frac{\left(1-\frac{1}{\alpha}\right)\sigma_a + \frac{Y}{\alpha}}{1-\xi-\frac{1}{1-\xi}\exp\left[\frac{\Delta}{b_{11}}\sigma_a + \frac{b_{11}-b_{12}}{b_{11}}\ln(1-\xi)+C\right]} \tag{16}$$

where $Y = \frac{2c\cos\phi}{1-\sin\phi}$; and

$$C = -\frac{\Delta}{b_{11}}\sigma_{r,ep} - \frac{b_{11}-b_{12}}{b_{11}}\ln(1-\xi_{ep}) \tag{17}$$

in which $\xi_{ep} = \frac{\sigma_{r,ep}-\sigma_h}{2G}$ pertains to the value of $\xi$ when the soil particle just starts yielding, i.e., at the stress state of $\mathbf{X_{ep}}$; and the corresponding radial stress $\sigma_{r,ep} = \sigma_h(1+\sin\phi) + c\cos\phi$.

Eq. (16) presents a simple first-order differential equation for the radial stress at the cavity wall/internal cavity pressure, $\sigma_a$, with $\xi$ being the independent variable. The equation is valid for any values of $\xi$ greater than $\xi_{ep}$ and up to 1, and can be readily solved as an initial value problem numerically so as to determine the variation of $\sigma_a$ with $\xi$, and subsequently the cavity pressure-expansion curve via the relationship $\xi = \frac{a-a_0}{a}$. The limiting cavity pressure $\sigma_u$ at $\mathbf{X_u}$ may be obtained by solving the same differential equation with $\xi \to 1$. Note that by substitution of $D\xi = \frac{1}{(a/a_0)^2} D(\frac{a}{a_0})$, Eq. (16) can be rewritten in the form

$$\frac{D\sigma_a}{D(a/a_0)} = -\frac{\left[\left(1-\frac{1}{\alpha}\right)\sigma_a+\frac{Y}{\alpha}\right]/(a/a_0)}{1-(a/a_0)^2 \exp\left[\frac{A}{b_{11}}\sigma_a - \frac{b_{11}-b_{12}}{b_{11}}\ln\left(\frac{a}{a_0}\right)+C\right]} \tag{18}$$

from which now the $\sigma_a - \frac{a}{a_0}$ relation/curve can be directly calculated while $\sigma_u$ may be obtained by putting $\frac{a}{a_0} \to \infty$.

**(b) Case II: $2\nu < 1 - \sin\phi$**

In this case under the category of $K_0 = 1$, the elastic stress path remains to be the same horizontal segment of $\mathbf{X_0 X_{ep}}$ as in Case I, see Fig. 1(b). The orientation of the incremental stress vector $D\boldsymbol{\sigma}_{ep}$ and the Lode angle of $D\mathbf{s}_{ep}$ are also still given by Eqs. (8) and (9). However, when $2\nu < 1 - \sin\phi$, the value of $\theta_{Ds_{ep}}$ according to Eq. (9) becomes greater than $\frac{11\pi}{6}$. This implies that the plastic stress path $\mathbf{X_{ep}X}$ may eventually intersect with the projected $s_r$ axis at certain point, say $\mathbf{X_{pr}}$, as shown in Fig. 1(b), unless the limit cavity pressure $\sigma_u$ (as $\xi \to 1$) has already been reached before such an intersection point ever been attained. If $\mathbf{X_{ep}X}$ does tend to meet with the $s_r$ axis after sufficient expansion of the cavity, then beyond point $\mathbf{X_{pr}}$, there are three possible stress paths that can be subsequently followed by the soil particle in compliance with continued loading on the cavity, i.e., $\mathbf{X_{pr}X'}$, $\mathbf{X_{pr}\overline{X}}$, and $\mathbf{X_{pr}\underline{X}}$ remaining on, deviating upward and downward from the line of $\theta = \frac{11\pi}{6}$, respectively (see Fig. 1(b)). However, as pointed out in Chen & Wang [23], the second scenario of $\mathbf{X_{pr}\overline{X}}$ is evidently invalid, since it contradicts with the downward movement trend featured by any possible stress path lying within the sextant of $\frac{3\pi}{2} \leq \theta \leq \frac{11\pi}{6}$ as interpreted in the previous subsection. On the other hand, if the stress path moves along the third direction of $\mathbf{X_{pr}X}$ beneath the line $\theta = \frac{11\pi}{6}$, it follows from the geometrical representation of the stress and strain states/increments [23] that the total deviatoric strain increment vector $D\mathbf{e}$ must also orient in a direction below the projected $s_r$ axis. Although not deemed necessary to be strictly parallel to the horizontal ($x$) axis as is demanded in the undrained case, such a resultant orientation of the $D\mathbf{e}$ vector nevertheless implies that $D\varepsilon_\theta > D\varepsilon_z = 0$. This is because the projected $D\varepsilon_r$ axis (on which $D\varepsilon_\theta = D\varepsilon_z$) actually represents a bisector line dividing the $\pi$-plane into two parts: the upper half corresponding to $D\varepsilon_\theta < D\varepsilon_z$ while the lower half to $D\varepsilon_\theta > D\varepsilon_z$, as illustrated in Fig. 1(b). Note that in the figure the three principal stresses and the incremental principal strains are plotted on the same axes [26]. The stress path $\mathbf{X_{pr}X}$ thus leads to a positive/compressive tangential strain increment, which is inconsistent with the well-known cavity expansion condition $D\varepsilon_\theta = -\frac{Dr}{r} < 0$.

The above graphical analysis proves that the only possible stress path following point $\mathbf{X_{pr}}$ must be $\mathbf{X_{pr}X'}$ along the projected $s_r$ axis, which is again analogous to the undrained case [23]. In this circumstance, the stress state $\mathbf{X'}$ maintains on an edge of the Mohr-Coulomb yield surface; the two active yield and potential functions contributing to the plastic flow are thus $F_{r\theta}$ and $P_{r\theta}$ as already defined in Eqs. (1) and (6), and those related to $\sigma_r$ and $\sigma_z$ in the sextant of $-\frac{\pi}{6} < \theta < \frac{\pi}{6}$:

$$F_{rz}(\sigma_r, \sigma_z) = \frac{\sigma_r - \sigma_z}{2} - \frac{\sigma_r + \sigma_z}{2}\sin\phi - c\cos\phi \tag{19}$$

$$P_{rz}(\sigma_r, \sigma_z) = \frac{\sigma_r - \sigma_z}{2} - \frac{\sigma_r + \sigma_z}{2}\sin\psi - c_{rz}^*\cos\psi \tag{20}$$

where $c_{rz}^*$ is another constant and only depends on $\mathbf{X_{pr}}$. Hence, the corresponding constitutive equation for this corner loading case can be expressed as [23]

$$\begin{Bmatrix} D\sigma_r \\ D\sigma_\theta \\ D\sigma_z \end{Bmatrix} = \frac{1}{\Delta_r} \begin{bmatrix} (1+m)(1+n) & (1+m)(1-n) & (1+m)(1-n) \\ (1-m)(1+n) & (1-m)(1-n) & (1-m)(1-n) \\ (1-m)(1+n) & (1-m)(1-n) & (1-m)(1-n) \end{bmatrix} \begin{Bmatrix} D\varepsilon_r \\ D\varepsilon_\theta \\ D\varepsilon_z \end{Bmatrix} \qquad (21)$$

where $m = \sin\phi$; $n = \sin\psi$; and $\Delta_r = \frac{1}{2G(1+\nu)}[3 - 6\nu + (2\nu - 1)(m + n) + (3 + 2\nu)mn]$.

As with Eq. (11), Eq. (21) provides only one single independent equation, as a result of $D\sigma_\theta = D\sigma_z = \frac{1}{\alpha}D\sigma_r$ that needs to be satisfied when the stress point $\mathbf{X}'$ moves along the line $\theta = \frac{11\pi}{6}$. Making again use of the first row of Eq. (21) and then in combination with the Lagrangian form of the radial equilibrium condition (13), one can readily obtain

$$\frac{D\sigma_a}{D\xi} = -\frac{\left(1-\frac{1}{\alpha}\right)\sigma_a + \frac{Y}{\alpha}}{1-\xi-\frac{1}{1-\xi}\exp\left[\frac{\Delta_r}{(1+m)(1+n)}\sigma_a + \frac{2n}{(1+n)}\ln(1-\xi) + C_r\right]} \qquad (22)$$

by following a similar procedure as described before for the case of $2\nu \geq 1 - \sin\phi$. The solution constant $C_r$ appearing in the above equation can be expressed as

$$C_r = \left[\frac{\Delta}{b_{11}} - \frac{\Delta_r}{(1+m)(1+n)}\right]\sigma_{r,pr} + \left[\frac{b_{11}-b_{12}}{b_{11}} - \frac{2n}{(1+n)}\right]\ln(1-\xi_{pr}) + C \qquad (23)$$

where $\sigma_{r,pr}$ denotes the radial stress pertaining to $\mathbf{X_{pr}}$ and $\xi_{pr}$ the corresponding value of $\xi$.

Note that the governing equation (22) is intended for the cavity responses analysis beyond the stress state $\mathbf{X_{pr}}$, and hence applies only to the situation of $\xi \geq \xi_{pr}$. Here $\xi_{pr}$ should be sought by solving inversely the differential equation (16) up to the critical stress state of $\sigma_r = \sigma_{r,pr}$, with $\sigma_{r,pr}$ being determined in a graphical manner as follows. Consider the triangle $\mathbf{X_0 X_{ep} X_{pr}}$ illustrated in Fig. 1(b) and recall Eq. (9), it can be easily found that the obtuse angle between $\mathbf{X_0 X_{ep}}$ and $\mathbf{X_{ep} X_{pr}}$ is equal to $\frac{2\pi}{3} - \tan^{-1}\left(\frac{1-3\sin\phi-2\nu}{\sqrt{3}[1-2\nu+\sin\phi]}\right)$. Now by the condition $|\mathbf{X_0 X_{ep}}| = \frac{\sqrt{2}}{\sqrt{3}}q_{ep} = \sqrt{2}(\sigma_\nu \sin\phi + c\cos\phi)$ (where $q_{ep}$ denotes the deviatoric stress at point $\mathbf{X_{ep}}$), $\mathbf{X_0 X_{pr}}$ hence is of length

$$|\mathbf{X_0X_{pr}}| = \sqrt{2}(\sigma_v \sin\phi + c\cos\phi) \frac{\sin\left\{\frac{2\pi}{3} - \tan^{-1}[(1-3\sin\phi-2\nu)/\sqrt{3}(1-2\nu+\sin\phi)]\right\}}{\sin\left\{\frac{\pi}{6} + \tan^{-1}[(1-3\sin\phi-2\nu)/\sqrt{3}(1-2\nu+\sin\phi)]\right\}} \quad (24)$$

and so

$$\sigma_{r,pr} = \frac{1}{1-\alpha}\left\{Y - \alpha\frac{\sqrt{3}}{\sqrt{2}}|\mathbf{X_0X_{pr}}|\right\}$$

$$= \frac{Y}{1-\alpha} - \frac{\sqrt{3}\alpha}{1-\alpha}(\sigma_v\sin\phi + c\cos\phi)\frac{\sin\left\{\frac{2\pi}{3} - \tan^{-1}[(1-3\sin\phi-2\nu)/\sqrt{3}(1-2\nu+\sin\phi)]\right\}}{\sin\left\{\frac{\pi}{6} + \tan^{-1}[(1-3\sin\phi-2\nu)/\sqrt{3}(1-2\nu+\sin\phi)]\right\}} \quad (25)$$

In summary, Eqs. (16) and (22) are the basis for the plastic solution of the cavity expansion problem in a non-associated Mohr-Coulomb soil, for the case of $K_0 = 1$. The former equation is applicable to the conditions $2\nu \geq 1 - \sin\phi$, as well as to $2\nu < 1 - \sin\phi$ provided that the internal cavity pressure $\sigma_a$ is not exceeding the critical value of $\sigma_{r,pr}$ as given by Eq. (25). In contrast, if both the two inequalities $2\nu < 1 - \sin\phi$ and $\sigma_a > \sigma_{r,pr}$ hold true, Eq. (22) nevertheless should be used in determining the cavity expansion curve and the limit cavity pressure. Finally, it is to be remarked that although the deviatoric stress paths developed under drained and undrained [23] conditions are exactly the same, this by no means implies that these two expansion cases will lead to identical strain/deformation responses of the cavity and/or pressure-expansion relation (they indeed must not).

## NON-HYDROSTATIC IN-SITU STRESS CONDITION WITH $K_0 < 1$

For the case of $K_0 < 1$, point $\mathbf{X_0}$, the projection of the in situ stress state on the $\pi$-plane, is now located on the triaxial compression line with $\theta = \frac{7\pi}{6}$ (or $y$ axis), as shown in Fig. 2. Consider further the relative position of $\mathbf{X_0}$ with respect to a base point $R$ such chosen that $RA$ parallels the horizontal ($x$) axis, two situations then arise: $\mathbf{X_0}$ lies either on the lower segment of $OR$ (Fig. 2(a)) or on the upper segment $RV$ (Figs. 2(b)). Let $K_{0(R)}$ denote the $K_0$ value at point $R$, which can be calculated as (see the Appendix)

$$K_{0(R)} = \frac{1-(c/\sigma_v)\cos\phi}{1+\sin\phi} \tag{26}$$

Fig. 2(a) and Figs. 2(b) therefore correspond to two respective cases of $K_{0(R)} < K_0 < 1$ and $K_{0(V)} < K_0 < K_{0(R)}$. Here

$$K_{0(V)} = \frac{1-\sin\phi-2(c/\sigma_v)\cos\phi}{1+\sin\phi} \tag{27}$$

which represents the $K_0$ value at the major vertex $V$. Note that Eq. (27) results directly from the application of the yielding condition to point $V$. For given Mohr-Coulomb parameters of $c$ and $\phi$, yielding will occur even under the in situ stress conditions if $K_0 = K_{0(V)}$.

**(c) Case III: $K_{0(R)} < K_0 < 1$**

During the elastic phase of the cylindrical cavity expansion, the following relationship $D\sigma_z = \frac{1}{2}(D\sigma_r + D\sigma_\theta) \equiv 0$ always holds among the three stress increments [6]. This implies that the resulting elastic stress path must be directed horizontally regardless of the magnitude of $K_0$. Now from the geometry of Fig. 2(a), it can be found that in the present case of $K_{0(R)} < K_0 < 1$, the end point of the horizontal elastic stress path, $\mathbf{X_{ep}}$, is still located in the sector of $\frac{3\pi}{2} < \theta < \frac{11\pi}{6}$. As a result, the yield and potential surfaces involved in the determination of the stress path within this sector and subsequently the analysis of the cavity responses will still be associated with the two functions of $F_{r\theta}$ and $P_{r\theta}$ given by Eqs. (1) and (6) as used for the case of $K_0 = 1$. The shaded area admissible to all the possible stress increments and the pattern of the resultant stress paths in the $\pi$-plane for the two scenarios of $2\nu \geq 1 - \sin\phi$ and $2\nu < 1 - \sin\phi$, as shown collectively in Fig. 2(a), therefore are not much different from those already presented in Figs. 1(a) and 1(b). If, under certain circumstances, the stress path eventually reaches the line of $\theta = \frac{11\pi}{6}$ at the intersection point $\mathbf{X_{pr}}$, it will again stick to that line afterwards until the ultimate state $\mathbf{X_u}$ is arrived at. The corresponding cavity responses analysis involving the two yield surfaces, $F_{r\theta}$ and $F_{rz}$, and two potential surfaces, $P_{r\theta}$ and $P_{rz}$, can again be carried out in the same way as the $K_0 = 1$ case. The above description clearly indicates that the cavity expansion solution in a Mohr-Coulomb soil, with more general values of $K_0$ less than 1 but greater than $K_{0(R)}$, is still obtainable from Eqs. (16) and (22).

**(d) Case IV: $K_{0(V)} < K_0 < K_{0(R)}$**

The solution for the case of $K_{0(V)} < K_0 < K_{0(R)}$, i.e., $\mathbf{X_0}$ located above the reference point $R$, can still be obtained by following the present graphical-analysis based method, although the stress path may become more complicated. As can be seen from Fig. 2(b), the main difference of the stress paths between this case and the previous ones lies in that the elastic-plastic transition stress point $\mathbf{X_{ep}}$ now shifts upwards to a new plane face of the Mohr-Coulomb yield surface, governed by

$$F_{z\theta}(\sigma_z, \sigma_\theta) = \frac{\sigma_z - \sigma_\theta}{2} - \frac{\sigma_z + \sigma_\theta}{2}\sin\phi - c\cos\phi \tag{28}$$

with the accompanying potential function

$$P_{z\theta}(\sigma_z, \sigma_\theta) = \frac{\sigma_z - \sigma_\theta}{2} - \frac{\sigma_z + \sigma_\theta}{2}\sin\psi - c^*_{z\theta}\cos\psi \tag{29}$$

where $c^*_{z\theta}$ is known as a constant. Accordingly, the developed plastic stress paths will in general span the two adjacent sextants of $\frac{7\pi}{6} < \theta < \frac{3\pi}{2}$ and $\frac{3\pi}{2} < \theta < \frac{11\pi}{6}$. Consider first the stress increment $D\mathbf{s}_{ep}$. Following a similar way as in the graphical proof of $\mathbf{X_{pr}\underline{X}}$ being inconsistent with the cavity expansion condition $D\varepsilon_\theta < 0$, it can be readily verified that $D\mathbf{s}_{ep}$, or the portion of the plastic stress path $\mathbf{X_{ep}X_{p\theta}}$, must be directed towards the negative $s_\theta$ axis ($\theta = \frac{3\pi}{2}$), as illustrated in Fig. 2(b). Here $\mathbf{X_{p\theta}}$ denotes the intersection point of the stress path with $s_\theta$ axis. Note that in arriving at such a conclusion, an additional characteristic feature of the cavity expansion problem that the radial strain increment $D\varepsilon_r$ is always compressive [28], and hence $D\varepsilon_r > 0 = D\varepsilon_z$, has been accounted for. It is also necessary to note that, unlike the known stress path of straight line within the sector of $\frac{3\pi}{2} < \theta < \frac{11\pi}{6}$ [see Eq. (8)], the trajectory $\mathbf{X_{ep}X_{p\theta}}$ laid out in the sextant $\frac{7\pi}{6} < \theta < \frac{3\pi}{2}$ is a curved one in general, which on its own needs to be determined from the equation of equilibrium in addition to the Mohr-Coulomb yield/constitutive conditions.

When the stress state $\widehat{\mathbf{X}}$ lies on the path $\mathbf{X_{ep}X_{p\theta}}$, the corresponding elastoplastic constitutive relation, analogous to the analysis for the case of $K_0 = 1$ with $2\nu \geq 1 - \sin\phi$, is given by

$$\begin{Bmatrix} D\sigma_r \\ D\sigma_\theta \\ D\sigma_z \end{Bmatrix} = \frac{1}{\Delta}\begin{bmatrix} b_{33} & b_{32} & b_{31} \\ b_{23} & b_{22} & b_{21} \\ b_{13} & b_{12} & b_{11} \end{bmatrix}\begin{Bmatrix} D\varepsilon_r \\ D\varepsilon_\theta \\ D\varepsilon_z \end{Bmatrix} \tag{30}$$

where $\Delta$ and $b_{ij}$ have been defined previously.

It should be remarked that from Eq. (30) the condition of $D\sigma_\theta = \frac{1}{\alpha}D\sigma_z$ is automatically satisfied, but there follows no such a relationship as Eq. (7) since, according to Eq. (29), the plastic flow in $z$ direction, $D\varepsilon_z^p$, is non-zero. Two independent equations thus ensue from the above stress-strain relations. Combining the first two of Eq. (30) once again with the Lagrangian form of the radial equilibrium equation (13), and after some arrangement, results in

$$\frac{D\sigma_a}{D\xi} = -\frac{\left(1 - \frac{b_{23}}{b_{33}}\right)\sigma_a - \frac{b_{33}b_{22} - b_{32}b_{23}}{\Delta b_{33}}\ln(1-\xi) - D_{z\theta}}{1 - \xi - \frac{1}{1-\xi}\exp\left[\frac{\Delta}{b_{33}}\sigma_a + \frac{b_{33} - b_{32}}{b_{33}}\ln(1-\xi) + C_{z\theta}\right]} \tag{31}$$

$$\sigma_{\theta,a} = \frac{b_{23}}{b_{33}}\sigma_a + \frac{b_{33}b_{22} - b_{32}b_{23}}{\Delta b_{33}}\ln(1-\xi) + D_{z\theta} \tag{32}$$

where

$$C_{z\theta} = -\frac{\Delta}{b_{33}}\sigma_{r,ep} - \frac{b_{33} - b_{32}}{b_{33}}\ln(1-\xi_{ep}) \tag{33a}$$

$$D_{z\theta} = \sigma_{\theta,ep} - \frac{b_{23}}{b_{33}}\sigma_{r,ep} - \frac{b_{33}b_{22} - b_{32}b_{23}}{\Delta b_{33}}\ln(1-\xi_{ep}) \tag{33b}$$

in which $\sigma_{r,ep} = (2K_0 - \frac{1-\sin\phi}{1+\sin\phi})\sigma_v + \frac{2c\cos\phi}{1+\sin\phi}$; $\sigma_{\theta,ep} = \frac{1-\sin\phi}{1+\sin\phi}\sigma_v - \frac{2c\cos\phi}{1+\sin\phi}$; and $\xi_{ep}$ has been defined after Eq. (17). It is clear that Eq. (31) can and should be solved for the variation of the cavity pressure up to $\sigma_a = \sigma_{r,p\theta}$ in terms of $\xi$, where $\sigma_{r,p\theta}$ is the radial stress pertaining to $\mathbf{X_{p\theta}}$. Note that $\sigma_{r,p\theta}$ itself has to be obtained from integrating the same differential equation (31) with the implementation of the following algebraic condition

$$\sigma_a = \frac{1}{1-\alpha(b_{23}/b_{33})}\left\{\alpha\left[\frac{b_{33}b_{22}-b_{32}b_{23}}{\Delta b_{33}}\ln(1-\xi) + D_{z\theta}\right] + \sigma_v - \alpha\sigma_{\theta,ep}\right\} \tag{34}$$

which is basically a restatement of the fact that the radial and vertical stresses are equal at the intersection point $\mathbf{X_{p\theta}}$. The assertion that the stress path $\mathbf{X_{ep}X_{p\theta}}$ should be curved now also becomes evident following Eq. (32), attributed to the appearance of the logarithmic term with respect to $\xi$ in the expression.

Next consider the cavity responses for the stress state proceeding beyond the point $\mathbf{X_{p\theta}}$. At this stage the stress path will enter into the sextant of $\frac{3\pi}{2} < \theta < \frac{11\pi}{6}$, and it may either be confined to this sector with $\sigma_r > \sigma_z > \sigma_\theta$ strictly valid throughout the expansion ($\mathbf{X_{p\theta}XX_u}$ in Figs. 2(b)) or end up with lying on the projected $s_r$ axis with $\sigma_\theta = \sigma_z$ ($\mathbf{X_{p\theta}XX_{pr}X'X_u}$ in the same figure), depending on the relative value of $2\nu$ with respect to $1 - \sin\phi$. This is indeed expected, since once the stress state has been brought to the sextant of $\frac{3\pi}{2} < \theta < \frac{11\pi}{6}$, the graphical analysis for the stress path and cavity responses given previously for the case of $K_{0(R)} < K_0 < 1$ will become applicable with no alteration to the present situation. With reference to Fig. 2(b), it therefore can be concluded that in the calculation of the stresses and deformations for the cavity problem, Eq. (31) should be used if the applied cavity pressure $\sigma_a \leq \sigma_{r,p\theta}$, no matter the values of $\nu$ and $\phi$; otherwise Eq. (16) is required for $2\nu \geq 1 - \sin\phi$, or for $2\nu < 1 - \sin\phi$ when $\sigma_a \leq \sigma_{r,pr}$ holds true; while Eq. (22) applies to the remaining case that the conditions of $2\nu < 1 - \sin\phi$ and $\sigma_a > \sigma_{r,pr}$ are both satisfied.

## NON-HYDROSTATIC IN-SITU STRESS CONDITION WITH $K_0 > 1$

The graphical analysis for this category of $K_0$ values is as the preceding one, except that the in situ stress point $\mathbf{X_0}$ is now on the triaxial tension line with $\theta = \frac{\pi}{6}$, so that the plastic stress path may go through the sector between the two lines of $\theta = \frac{11\pi}{6}$ (or $-\frac{\pi}{6}$) and $\theta = \frac{\pi}{6}$ before approaching the $s_r$ axis (Fig. 3(b)). Similar to the previous solution of $K_0 < 1$, the stress path and cavity responses analysis will be presented for the following two separate scenarios of $1 < K_0 < K_{0(S)}$ and $K_{0(S)} < K_0 < K_{0(T)}$, depending again on the position of $\mathbf{X_0}$ relative to another base point $S$ such that $SB \parallel Ox$, as shown in Fig. 3. Note that the $K_0$ value at point $S$ is given by (see the Appendix)

$$K_{0(S)} = \frac{1+(c/\sigma_v)\cos\phi}{1-\sin\phi} \tag{35}$$

and at the minor vertex $T$ by

$$K_{0(T)} = \frac{1+\sin\phi+2(c/\sigma_v)\cos\phi}{1-\sin\phi} \tag{36}$$

Here $K_{0(T)}$ corresponds to the maximum admissible value of $K_0$, below which the initial stress state of the soil, as required, will remain elastic and lie within the yield surface.

**(e) Case V: $1 < K_0 < K_{0(S)}$**

In this case $\mathbf{X_0}$ lies on the upper segment of $OS$, and according to Fig. 3(a), the transition point $\mathbf{X_{ep}}$ turns out to be located again in the favourable sector of $\frac{3\pi}{2} < \theta < \frac{11\pi}{6}$. As such, the governing equations (16) and (22) derived earlier for the conditions of $K_{0(R)} < K_0 \leq 1$ must still be valid for the current case. The cavity responses hence can be easily determined by solving these two differential equations.

**(f) Case VI: $K_{0(S)} < K_0 < K_{0(T)}$**

The solution for this last case is analogous to the previous one corresponding to $K_{0(V)} < K_0 < K_{0(R)}$. Note that now the transition stress point $\mathbf{X_{ep}}$, as shown in Fig. 3(b), has moved downward and been situated on the face of the Mohr-Coulomb yield surface defined by Eq. (19), the non-associated condition being Eq. (20). As before, the elastoplastic constitutive equation for stress state $\mathbf{X_{ep}}$ or any other state $\tilde{\mathbf{X}}$ lying on the path $\mathbf{X_{ep}X_{pr}}$ (exclusive of point $\mathbf{X_{pr}}$) can be written as

$$\begin{Bmatrix} D\sigma_r \\ D\sigma_\theta \\ D\sigma_z \end{Bmatrix} = \frac{1}{\Delta}\begin{bmatrix} b_{11} & b_{13} & b_{12} \\ b_{31} & b_{33} & b_{32} \\ b_{21} & b_{23} & b_{22} \end{bmatrix}\begin{Bmatrix} D\varepsilon_r \\ D\varepsilon_\theta \\ D\varepsilon_z \end{Bmatrix} \tag{37}$$

Again, keep in mind that the stress path $\mathbf{X_{ep}X_{pr}}$ followed in the sector $-\frac{\pi}{6} < \theta < \frac{\pi}{6}$ will be a curved line and cannot be directly determined graphically without considering the equilibrium equation. As with Eq. (30), Eq. (37) still gives two independent equations, which, along with the converted form of radial equilibrium condition (13), leads to similar differential/algebraic equations as Eqs. (31) and (32) for the determination of radial and tangential stresses at the cavity wall

$$\frac{D\sigma_a}{D\xi} = -\frac{\left(1-\frac{b_{31}}{b_{11}}\right)\sigma_a - \frac{b_{11}b_{33}-b_{13}b_{31}}{\Delta b_{11}}\ln(1-\xi)-D_{rz}}{1-\xi-\frac{1}{1-\xi}\exp\left[\frac{\Delta}{b_{11}}\sigma_a+\frac{b_{11}-b_{13}}{b_{11}}\ln(1-\xi)+C_{rz}\right]} \tag{38}$$

$$\sigma_{\theta,a} = \frac{b_{31}}{b_{11}}\sigma_a + \frac{b_{11}b_{33}-b_{13}b_{31}}{\Delta b_{11}}\ln(1-\xi) + D_{rz} \tag{39}$$

where

$$C_{rz} = -\frac{\Delta}{b_{11}}\sigma_{r,ep} - \frac{b_{11}-b_{13}}{b_{11}}\ln(1-\xi_{ep}) \tag{40a}$$

$$D_{rz} = \sigma_{\theta,ep} - \frac{b_{31}}{b_{11}}\sigma_{r,ep} - \frac{b_{11}b_{33}-b_{13}b_{31}}{\Delta b_{11}}\ln(1-\xi_{ep}) \tag{40b}$$

with $\sigma_{r,ep} = \frac{1+\sin\phi}{1-\sin\phi}\sigma_v + \frac{2c\cos\phi}{1-\sin\phi}$ and $\sigma_{\theta,ep} = \left(2K_0 - \frac{1+\sin\phi}{1-\sin\phi}\right)\sigma_v - \frac{2c\cos\phi}{1-\sin\phi}$.

It may be checked that, as long as the stress state $\tilde{\mathbf{X}}$ falls within the sector characterized by $\sigma_r > \sigma_\theta > \sigma_z$ so that the plastic straining is attributed solely to the single potential function defined by Eq. (20), the stress path $\mathbf{X_{ep}\tilde{X}}$ will always bend towards the projected $s_r$ axis, and tend to approach the intersection point $\mathbf{X_{pr}}$ at sufficiently expanded cavity radius $\frac{a}{a_0}$. This conclusion actually can be directly drawn, for the same reasoning that has led to the previous verification, for the case of $K_0 = 1$ with $2\nu < 1 - \sin\phi$, that $\mathbf{X_{pr}\underline{X}}$ below the line of $\theta = \frac{11\pi}{6}$ must not be a real possible stress path to follow.

Obviously Eq. (38) is valid for the cavity pressure only up to $\sigma_a = \sigma_{r,pr}$. Here $\sigma_{r,pr}$, however, is not supposed to be calculated from Eq. (25) (applicable only for Case II: $2\nu < 1 - \sin\phi$ with a straight stress path in the sextant $\frac{3\pi}{2} < \theta < \frac{11\pi}{6}$). Instead, in analogy with $\sigma_{r,p\theta}$ in Case IV: $K_{0(V)} < K_0 < K_{0(R)}$, it needs to be determined by solving the same differential equation (38) with the supplement of the equality condition between the tangential and vertical stresses at point $\mathbf{X_{pr}}$, given by

$$\sigma_a = \frac{1}{1/\alpha - b_{31}/b_{11}} \left[ \frac{b_{11}b_{33} - b_{13}b_{31}}{\Delta b_{11}} \ln(1-\xi) + D_{rz} - \sigma_v + \frac{1}{\alpha}\sigma_{r,ep} \right] \quad (41)$$

As $\sigma_a$ increases further, the stress path beyond point $\mathbf{X_{pr}}$ will again be controlled by the values of $2\nu$ and $1 - \sin\phi$. Fig. 3(b) illustrates the two typical stress paths that correspond to $2\nu \geq 1 - \sin\phi$ and $2\nu < 1 - \sin\phi$, respectively. The former of these two remains strictly in the sector of $\frac{3\pi}{2} < \theta < \frac{11\pi}{6}$ (bounded actually by $x$ and projected $s_r$ axes), for which Eq. (16) should be invoked in the analysis of the cavity responses; while the latter one continues staying along the projected $s_r$ axis and one hence needs to resort to Eq. (22) to obtain the cavity stress and deformation results.

At this stage, a complete set of analytical solutions for the cavity expansion in non-associated Mohr-Coulomb soil has been fully developed within the proposed graphical framework. There is however one subtle point that deserves more discussions. Recall that in dealing with Case IV:

$K_{0(V)} < K_0 < K_{0(R)}$ under the category of $K_0 < 1$, the stress path $\mathbf{X_{p\theta}X}$ beyond point $\mathbf{X_{p\theta}}$ has been defaulted in the sextant of $\frac{3\pi}{2} < \theta < \frac{11\pi}{6}$. Nevertheless geometrically it can also remain on the negative $s_\theta$ axis without violating any potential cavity deformation restraints, i.e., $\theta = \frac{3\pi}{2}$ with $D\sigma_r = D\sigma_z$, which therefore leads to an alternative solution provided that a somewhat similar constitutive relation as Eq. (21) is properly used in the analysis of cavity expansion afterwards. In other words, there may be a bifurcation of the cavity responses being triggered when the stress state reaches the line $\theta = \frac{3\pi}{2}$. However, according to the above formulations for the case of $K_{0(R)} < K_0 < 1$, it is apparent that for any stress state located infinitesimally below the negative $s_\theta$ axis, the subsequent stress path tends to show significant departure from the line of $\theta = \frac{3\pi}{2}$ and, indeed, has a Lode angle no less than $\frac{5\pi}{3}$. Considering that the stress path development/cavity responses are supposedly continuous with the stress state $\mathbf{X_{p\theta}}$ across the $s_\theta$ axis, the postulate of stress path $\mathbf{X_{p\theta}X}$ lying within the sector of $\frac{3\pi}{2} < \theta < \frac{11\pi}{6}$, is therefore deemed to be a reasonable one.

**Numerical results and discussions**

A numerical example is presented in this section to demonstrate the applications of the proposed graphical procedure for the cavity responses analysis. The new graphical method developed will first be compared with the seminal analytical (series) solution of Yu & Houlsby [3] that can be covered as the special Case I: $2\nu \geq 1 - \sin\phi$ under the category of $K_0 = 1$, i.e., the vertical stress strictly remains to be intermediate during the course of cavity expansion. Fig. 4 gives a comparison of the calculated cavity expansion curves for the following specified parameters [3]: $\nu = 0.3$; $\phi = 30°$; $\frac{2G(1+\nu)}{p_0+[Y/(\alpha-1)]} = 260$; and $\psi = 0°, 10°, 30°$. The excellent agreement observed between the two results clearly indicates the validity and accuracy of the presently proposed graphical solution procedure.

The (normalized) soil properties used in the parametric numerical analyses are: shear modulus $\frac{G}{\sigma_v} = 100$; Poisson's ratio $\nu = 0.3$; cohesion $\frac{c}{\sigma_v} = 0.5$; friction angle $\phi = 15°, 30°,$ and $45°$; and dilation angle $\psi = 10°$. Table 1 lists the four respective reference $K_0$ values, i.e., $K_{0(V)}$, $K_{0(R)}$, $K_{0(S)}$, and $K_{0(T)}$, related to the above three values of friction angle considered. To include a wide range of possible earth pressure coefficient at rest and to make full coverage of the various solution cases described in the preceding section, the following constant $K_0$ values of 0.3, 0.8, 1 and 1.5 are taken for all the three magnitudes of $\phi$ varied from 15° to 45° for Case I: $2\nu \geq 1 - \sin\phi$ through Case V: $1 < K_0 < K_{0(S)}$. Nevertheless, since there exists no single value of $K_0$ satisfying $K_{0(S)} < K_0 < K_{0(T)}$ for all the $\phi$ cases involved (see Table 1), different values of $K_0 = 2.5, 4,$ and 6 will be used in the numerical analysis for $\phi = 15°, 30°,$ and $45°$, respectively, so as to examine Case VI: $K_{0(S)} < K_0 < K_{0(T)}$ under the category of $K_0 > 1$ adequately.

Fig. 5 shows the influences of friction angle $\phi$ on the variation of normalized internal cavity pressure $\frac{\sigma_a}{\sigma_v}$ with the expanding cavity radius $\frac{a}{a_0}$, for a soil with an initial isotropic stress state ($K_0 = 1$). Note that the two cavity expansion curves of $\phi = 30°$ and $45°$ correspond to the solution Case I: $2\nu \geq 1 - \sin\phi$, while the curve of $\phi = 15°$ corresponds to Case II: $2\nu < 1 - \sin\phi$. Therefore, the possible intersection point $\mathbf{X_{pr}}$ with the projected $s_r$ axis may only occur for the latter case of $\phi = 15°$, as marked in Fig. 5. It is also unsurprising to see that the friction angle has a profound impact on the calculated cavity responses; an increased value of $\phi$ results in stiffer cavity expansion curves and hence higher limit cavity pressure $\sigma_u$.

Figs. 6-9 present the cavity expansion curves associated with the non-isotropic in situ stress cases of $K_0 = 0.8, 1.5, 0.3$ and $K_0 = 2.5, 4, 6$, again for all the three values of fiction angle

involved. As can be seen from the figure, for a given value of $\phi$, the calculated internal cavity pressure $\frac{\sigma_a}{\sigma_v}$ increases monotonously as $K_0$ increases from 0.3 to 6, which is indeed expected. Also, it appears that $\frac{a_{pr}}{a_0}$, the cavity radius pertaining to the intersection stress state $\mathbf{X_{pr}}$, tends to decrease with increasing $K_0$. For instance, with $\phi = 15°$, a larger value of $K_0 = 2.5$ yields a quite small value of $\frac{a_{pr}}{a_0} = 1.006$ (Fig. 9), as it compares with $\frac{a_{pr}}{a_0} = 1.08$ for the case of $K_0 = 1.5$ (Fig. 7) and $\frac{a_{pr}}{a_0} = 2.46$ for the case of $K_0 = 1$ (Fig. 5). As $K_0$ drops to 0.8 and 0.3 as shown in Figs. 6 and 8, the intersection point $\mathbf{X_{pr}}$ actually become unattainable (equivalent to $\frac{a_{pr}}{a_0} \to \infty$), indicating that in these two cases the limit cavity pressure $\sigma_u$ essentially has already been reached in the main sextant $\frac{3\pi}{2} \leq \theta \leq \frac{11\pi}{6}$ before the stress path hitting the projected $s_r$ axis.

The results for the limit cavity pressure $\sigma_u$ in variation with the soil friction angle $\phi$, for four different values of $K_0 = 0.3, 0.8, 1$, and 1.5, are further given in Fig. 10. Note that the figure has not included the curves pertinent to $K_0 = 2.5, 4$, and 6. This follows from the consideration that for each of these $K_0$ values, different/inconsistent solution procedures corresponding to Case V: $1 < K_0 < K_{0(S)}$ and Case VI: $K_{0(S)} < K_0 < K_{0(T)}$ have to be applied as $\phi$ varies from 10° to 50° (note: $K_{0(T)}$ calculated from Eq. (36) must not be less than $K_0$ if a valid $\phi$ is to be considered). Once again, it is clearly shown that the larger the in situ earth pressure coefficient, the higher is the value of $\sigma_u$ that can be reached. Like the undrained situation [23], the limit cavity pressure is observed to increase steadily and very near to linearly with the friction angle ranged from 10° to 50°, despite the fact that the solution procedure may have switched from solving Eq. (16) to Eq. (22) across the threshold value of $\phi_{th} = 23.58°$ (at which the critical condition $2\nu = 1 - \sin\phi$ is fulfilled).

**Conclusions**

A complete graphical solution procedure is rigorously developed for the first time, for the cylindrical cavity expansion problem in non-associated Mohr-Coulomb soil under drained conditions. The mathematical challenges encountered in the calculation of flow rate when the stress state lies on the corner/edge of two adjacent yield surfaces have been successfully tackled, with the aid of a generalized Koiter's theory for non-associated plasticity. The stress path to be followed by a soil particle during the cavity expansion can be determined/tracked through a unique geometrical analysis, on leveraging the deformation characteristic of the drained expansion problem that the corresponding radial and tangential strain increments must be compressive and tensile, respectively. The novel graph-based solution proposed circumvents the use of the traditional cumbersome zoning method involving the sequential determination of different Mohr-Coulomb plastic regions (with distinct principal stress orders) that may develop surrounding the cavity. More importantly, it is of sufficient generality to take into consideration of the effect of the vertical stress on the cylindrical cavity responses and to cover any arbitrary values of the in situ earth pressure coefficient as well.

It is found that the deviatoric stress path within the sextant of $\frac{3\pi}{2} < \theta < \frac{11\pi}{6}$ always remains to be a straight line as with the undrained expansion case. Nevertheless, outside of this major sector when the vertical stress becomes no longer the intermediate principal stress, the stress paths turn out to be curved ones so as to accommodate the generation of non-zero plastic flow in the vertical direction. The desired stress path and the cavity responses in general need to be determined through solving a single first-order differential equation with the internal cavity pressure and an introduced auxiliary variable being the basic unknowns. Some selected numerical results with various friction angle and in situ earth pressure coefficient reveal that the stiffness of the cavity expansion curve as well as the limit cavity pressure both increase profoundly with the increased values of these two essential soil parameters. The rigorous analytical solution proposed provides a completion of the drained analysis of cavity expansion problem with the classical Mohr-Coulomb model. It is deemed to be not only of great significance for more accurate interpretation of in-situ test results pertaining to cohesive-frictional soils with non-isotropic initial stress state, but also able to serve as a unique benchmark solution for truly verifying the correctness and capability of the cornered Mohr-Coulomb constitutive model built in commercial finite element programs.

**Funding statement**

This research is funded by the Industrial Ties Research Subprogram of the Louisiana Board of Regents [Grant No. LEQSF(2019-22)-RD-B-01].

**Appendix: Determination of $K_{0(R)}$ and $K_{0(S)}$**

Refer to Fig. 2(a), the in situ horizontal and vertical stresses of point $R$ can be expressed as [8, 11]

$$\sigma_{h,R} = \left(1 + 2K_{0(R)}\right)\sigma_{v,R} + \frac{2}{3}\left(\frac{\sqrt{3}}{\sqrt{2}}|OR|\right)\sin\left(\frac{7\pi}{6} + \frac{2\pi}{3}\right) = \left(1 + 2K_{0(R)}\right)\sigma_{v,R} - \frac{1}{\sqrt{6}}|OR| \quad (42a)$$

$$\sigma_{v,R} = \left(1 + 2K_{0(R)}\right)\sigma_{v,R} + \frac{2}{3}\left(\frac{\sqrt{3}}{\sqrt{2}}|OR|\right)\sin\left(\frac{7\pi}{6} - \frac{2\pi}{3}\right) = \left(1 + 2K_{0(R)}\right)\sigma_{v,R} + \frac{\sqrt{2}}{\sqrt{3}}|OR| \quad (42b)$$

Note that

$$|OR| = \frac{1}{2}|OA| = \left\{\frac{c}{\tan\phi} + \left(1 + 2K_{0(R)}\right)\sigma_{v,R}\right\}\frac{\sqrt{6}\sin\phi}{3+\sin\phi} \quad (43)$$

Substituting Eq. (43) into Eqs. (42a) and (42b), one can easily derive

$$K_{0(R)} = \frac{\sigma_{h,R}}{\sigma_{v,R}} = \frac{1-(c/\sigma_v)\cos\phi}{1+\sin\phi} \quad (44)$$

Similarly, for reference point $S$ (see Fig. 3(a)), the following two expressions can be obtained

$$\sigma_{h,S} = \left(1 + 2K_{0(S)}\right)\sigma_{v,S} + \frac{2}{3}\left(\frac{\sqrt{3}}{\sqrt{2}}|OS|\right)\sin\left(\frac{\pi}{6} + \frac{2\pi}{3}\right) = \left(1 + 2K_{0(S)}\right)\sigma_{v,S} + \frac{1}{\sqrt{6}}|OS| \quad (45a)$$

$$\sigma_{v,S} = \left(1 + 2K_{0(S)}\right)\sigma_{v,S} + \frac{2}{3}\left(\frac{\sqrt{3}}{\sqrt{2}}|OR|\right)\sin\left(\frac{\pi}{6} - \frac{2\pi}{3}\right) = \left(1 + 2K_{0(R)}\right)\sigma_{v,R} - \frac{\sqrt{2}}{\sqrt{3}}|OS| \quad (45b)$$

while

$$|OS| = \frac{1}{2}|OB| = \left\{\frac{c}{\tan\phi} + \left(1 + 2K_{0(S)}\right)\sigma_{v,S}\right\}\frac{\sqrt{6}\sin\phi}{3-\sin\phi} \quad (46)$$

so

$$K_{0(S)} = \frac{\sigma_{h,S}}{\sigma_{v,S}} = \frac{1 + (c/\sigma_v)\cos\phi}{1 - \sin\phi} \quad (47)$$

**Notation**

| | |
|---:|---|
| $a$ | current radius of (expanded) cylindrical cavity |
| $a_0$ | initial radius of cylindrical cavity |
| $b_{ij}$ | constant elements of elastoplastic constitutive matrix $(i,j = 1, 2, 3)$ |
| $C, C_r, C_{rz}, C_{z\theta}$ | solution constants |
| $c$ | cohesion |
| $c_{rz}^*, c_{r\theta}^*, c_{z\theta}^*$ | constant parameters involved in potential functions |
| $D_{rz}, D_{z\theta}$ | solution constants |
| $D\mathbf{e}$ | deviator of incremental strain vector |
| $Dp_{ep}$ | mean stress increment at point $\mathbf{X}_{ep}$ |
| $D\mathbf{s}$ | deviatoric stress increment vector |
| $D\mathbf{s}_{ep}$ | deviatoric stress increment vector at stress state $\mathbf{X}_{ep}$ |
| $D\varepsilon_r, D\varepsilon_\theta, D\varepsilon_z$ | total strain increments of a given material particle in radial, tangential, and vertical directions |
| $D\varepsilon_v$ | volumetric strain increment |
| $D\varepsilon_z^e, D\varepsilon_z^p$ | elastic and plastic strain increments of a given material particle in vertical direction |
| $D\boldsymbol{\sigma}_{ep}$ | incremental stress vector at stress state $\mathbf{X_{ep}}$ |
| $D\sigma_r, D\sigma_\theta, D\sigma_z$ | stress increments of a given material particle in radial, tangential, and vertical directions |
| $D\sigma_{r,ep}, D\sigma_{\theta,ep}$ | radial and tangential stress increments at transition state $\mathbf{X_{ep}}$ |
| $D\sigma_{z,ep}$ | vertical stress increment at transition state $\mathbf{X_{ep}}$ |
| $F_{rz}, F_{r\theta}, F_{z\theta}$ | yield functions of Mohr-Coulomb model |
| $G$ | shear modulus |
| $K_0$ | earth pressure coefficient at rest |
| $K_{0(R)}, K_{0(S)}$ | $K_0$ values corresponding to base points $R$ and $S$ |
| $K_{0(V)}, K_{0(T)}$ | $K_0$ values corresponding to major/minor vertices $V$ and $T$ |
| $m$ | material constant related to friction angle |
| $n$ | material constant related to dilation angle |
| $p_0$ | hydrostatic in situ stress |
| $P_{rz}, P_{r\theta}, P_{z\theta}$ | potential functions |
| $q_{ep}$ | deviatoric stress at point $\mathbf{X_{ep}}$ |
| $s_r, s_\theta, s_z$ | stress deviator components in radial, tangential, and vertical principal directions |
| $\mathbf{X}$ | current stress state projected on sextant $\frac{3\pi}{2} < \theta < \frac{11\pi}{6}$ of the $\pi$-plane |
| $\mathbf{X'}$ | current stress state projected on $s_r$ axis |
| $\widehat{\mathbf{X}}$ | current stress state projected on sextant $\frac{7\pi}{6} < \theta < \frac{3\pi}{2}$ of the $\pi$-plane |
| $\tilde{\mathbf{X}}$ | current stress state projected on sextant $-\frac{\pi}{6} < \theta < \frac{\pi}{6}$ of the $\pi$-plane |
| $\overline{\mathbf{X}}, \underline{\mathbf{X}}$ | stress state above and below the projected $s_r$ axis |
| $\mathbf{X_0}$ | in situ stress state projected on the $\pi$-plane |
| $\mathbf{X_{ep}}$ | elastic-plastic transition stress state in the deviatoric plane |
| $\mathbf{X_{pr}}$ | intersection point of plastic stress path with the projected $s_r$ axis |
| $\mathbf{X_{p\theta}}$ | intersection point of plastic stress path with the negative $s_\theta$ axis in deviatoric plane |
| $\mathbf{X_u}$ | ultimate stress state in the deviatoric plane |
| $\alpha$ | material constant |
| $\varDelta$ | constant parameter involved in the elastoplastic stiffness matrix |
| $\varDelta_r$ | constant parameter involved in the elastoplastic stiffness matrix pertaining to corner loading case |
| $\varepsilon_{v,a}$ | volumetric strain at cavity wall |
| $\theta$ | Lode angle |
| $\theta_{D\mathbf{s}}$ | Lode angle of $D\mathbf{s}$ |
| $\theta_{D\mathbf{s}_{ep}}$ | Lode angle of $D\mathbf{s}_{ep}$ |
| $\theta_{\mathbf{X_{ep}}B}$ | Lode angle of line $\mathbf{X_{ep}}B$ |
| $\lambda_{r\theta}$ | plastic multiplier corresponding to potential function $P_{r\theta}$ |
| $\nu$ | Poisson's ratio |
| $\xi$ | auxiliary variable |
| $\xi_{ep}$ | value of $\xi$ at point $\mathbf{X_{ep}}$ |
| $\sigma_a$ | internal cavity pressure |
| $\sigma_h$ | horizontal in situ stress |
| $\sigma_r, \sigma_\theta, \sigma_z$ | stress components in radial, tangential, and vertical directions |
| $\sigma_{r,ep}, \sigma_{\theta,ep}$ | radial and tangential stresses at point $\mathbf{X_{ep}}$ |
| $\sigma_{r,pr}$ | radial stress corresponding to intersection point $\mathbf{X_{pr}}$ |
| $\sigma_{r,p\theta}$ | radial stress corresponding to intersection point $\mathbf{X_{p\theta}}$ |
| $\sigma_u$ | limit cavity pressure |
| $\sigma_v$ | vertical in situ stress |
| $\sigma_{\theta,a}$ | tangential stress at cavity wall |
| $\phi$ | friction angle |
| $\phi_{th}$ | threshold value of friction angle |
| $\psi$ | dilation angle |

**Captions of tables and figures**

Table 1. $K_0$ values for reference points V, R, S, and T ($c/\sigma_v = 0.5$)

Fig. 1. Graphical representation of stress state/path in deviatoric plane for a soil element during cavity expansion process with $K_0 = 1$: (a) $2\nu \geq 1 - \sin\phi$; (b) $2\nu < 1 - \sin\phi$

Fig. 2. Graphical representation of stress state/paths in deviatoric plane for a soil element during cavity expansion process with $K_0 < 1$: (a) $K_{0(R)} < K_0 < 1$; (b) $K_{0(V)} < K_0 < K_{0(R)}$

Fig. 3. Graphical representation of stress state/paths in deviatoric plane for a soil element during cavity expansion process with $K_0 > 1$: (a) $1 < K_0 < K_{0(S)}$; (b) $K_{0(S)} < K_0 < K_{0(T)}$

Fig. 4. Comparisons of cavity expansion curves between current graphical method and Yu & Houlsby (1991) for the special case of $2\nu > 1 - \sin\phi$ with $K_0 = 1$

Fig. 5. Influences of friction angle on the variation of normalized internal cavity pressure with expansion ratio, $K_0 = 1$

Fig. 6. Influences of friction angle on the variation of normalized internal cavity pressure with expansion ratio, $K_0 = 0.8$

Fig. 7. Influences of friction angle on the variation of normalized internal cavity pressure with expansion ratio, $K_0 = 1.5$

Fig. 8. Influences of friction angle on the variation of normalized internal cavity pressure with expansion ratio, $K_0 = 0.3$

Fig. 9. Influences of friction angle on the variation of normalized internal cavity pressure with expansion ratio, $K_0 = 2.5$; $K_0 = 4$; and $K_0 = 6$

Fig. 10. Limit cavity pressure in variation with friction angle for various $K_0$ values

Table 1.  $K_0$ values for reference points V, R, S, and T ($c/\sigma_v = 0.5$)

| $\phi$ | $K_{0(V)}$ | $K_{0(R)}$ | $K_{0(S)}$ | $K_{0(T)}$ |
|---|---|---|---|---|
| 15° | −0.18 | 0.41 | 2.00 | 3.00 |
| 30° | −0.24 | 0.38 | 2.87 | 4.73 |
| 45° | −0.24 | 0.38 | 4.62 | 8.24 |

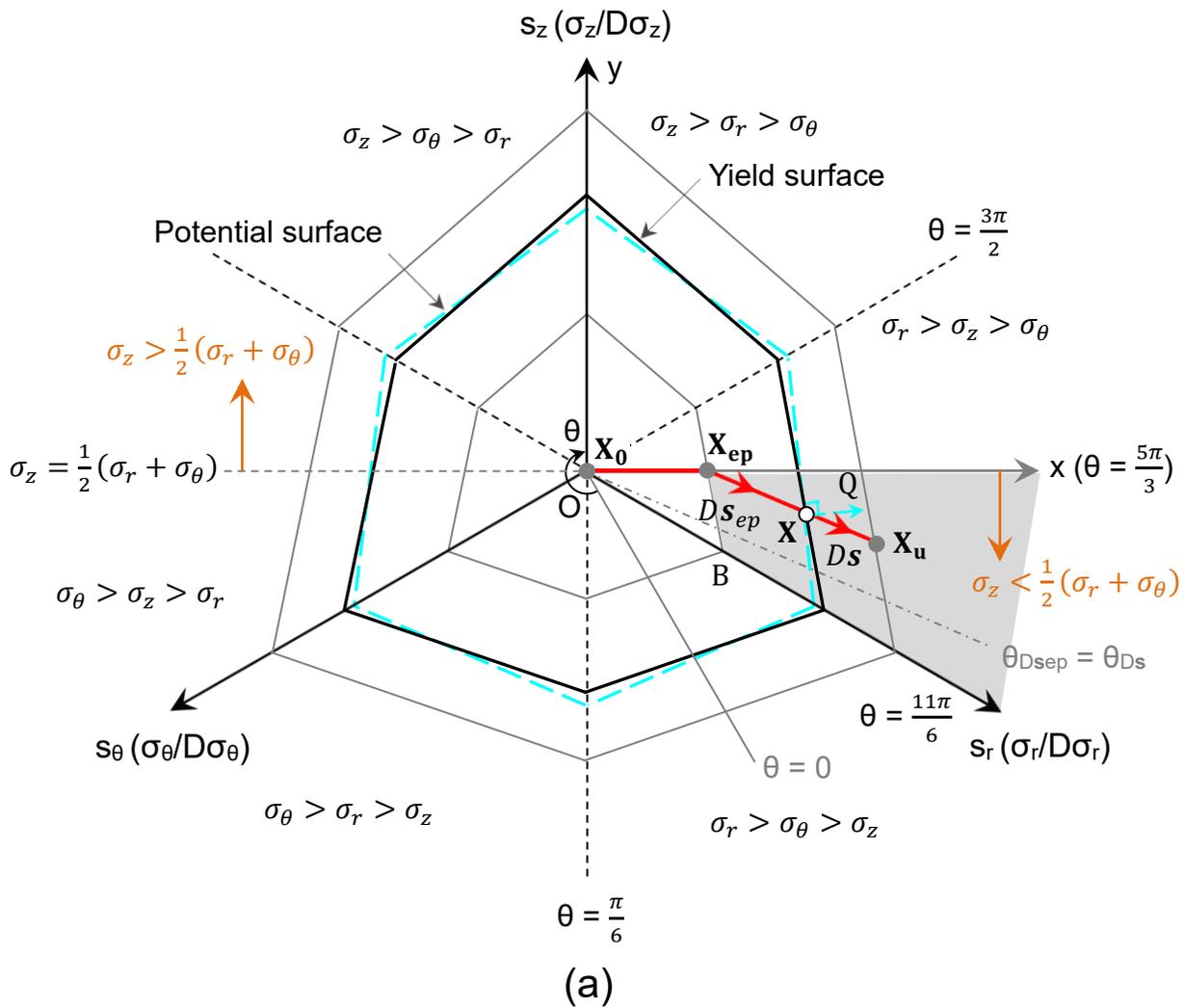

**Fig. 1.** Graphical representation of stress state/path in deviatoric plane for a soil element during cavity expansion process with $K_0 = 1$: (a) $2\nu \geq 1 - \sin\phi$; (b) $2\nu < 1 - \sin\phi$

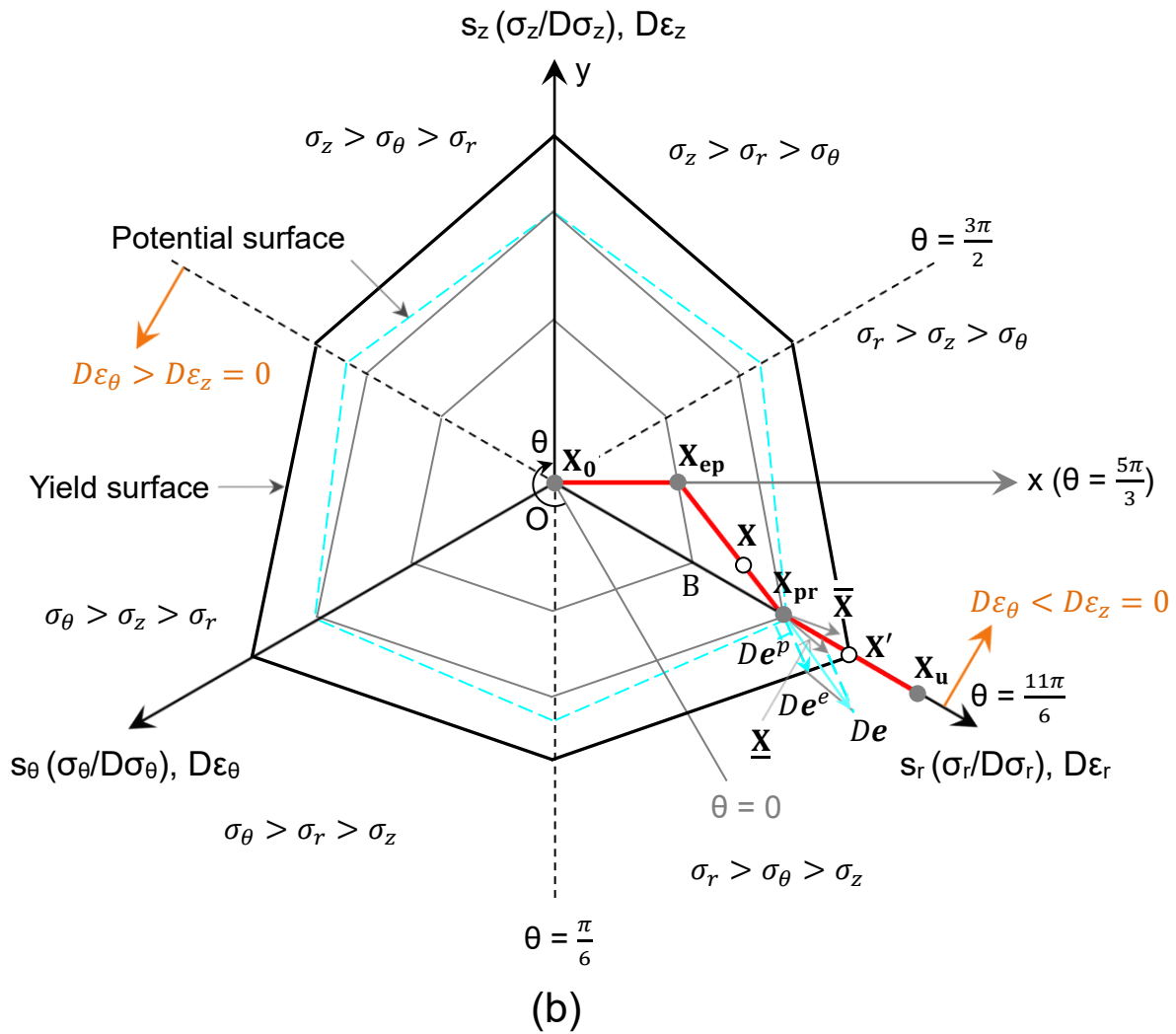

Fig. 1. (Cont'd) Graphical representation of stress state/path in deviatoric plane for a soil element during cavity expansion process with $K_0 = 1$: (a) $2\nu \geq 1 - \sin\phi$; (b) $2\nu < 1 - \sin\phi$

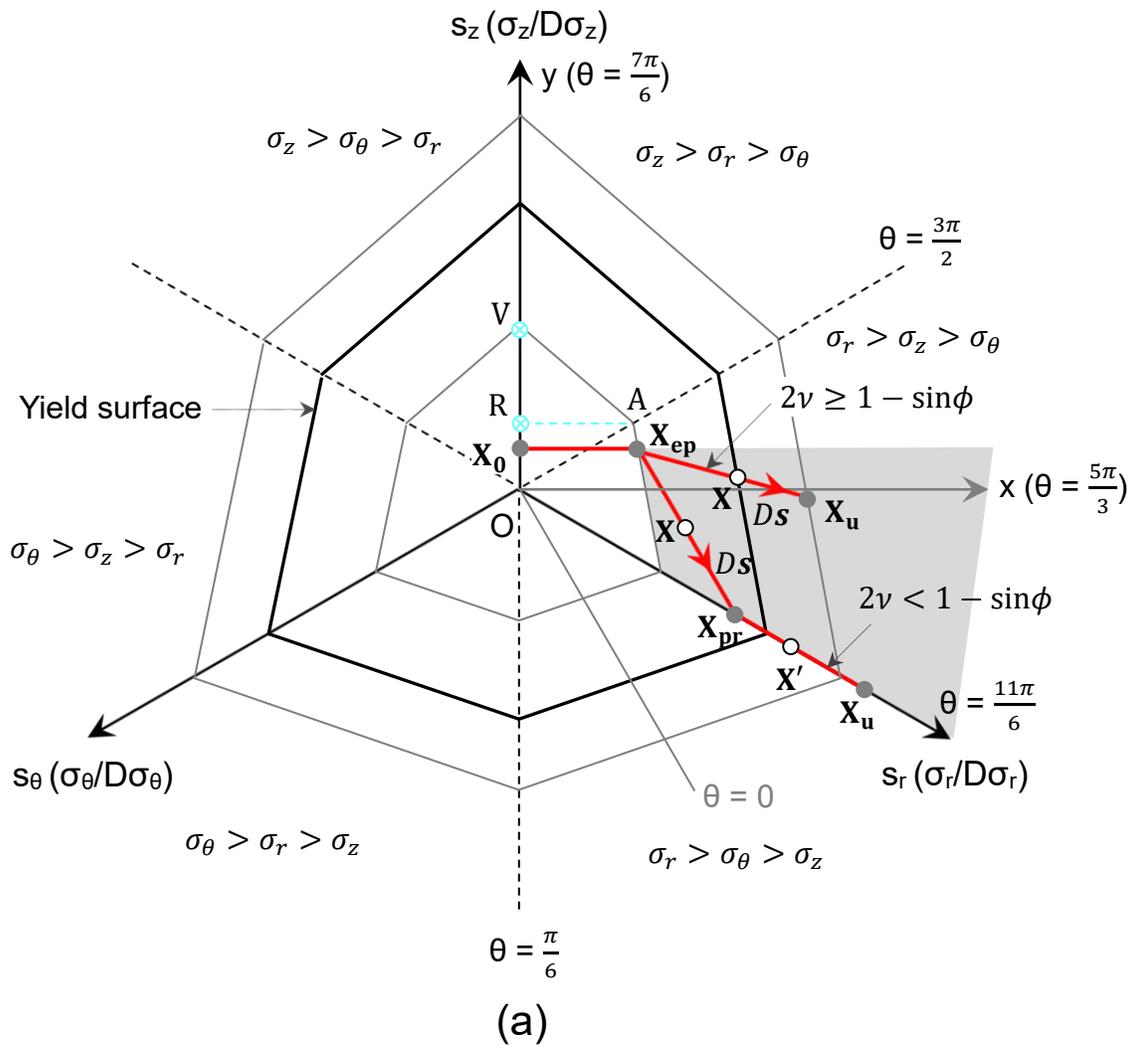

**Fig. 2.** Graphical representation of stress state/paths in deviatoric plane for a soil element during cavity expansion process with $K_0 < 1$: (a) $K_{0(R)} < K_0 < 1$; (b) $K_{0(V)} < K_0 < K_{0(R)}$

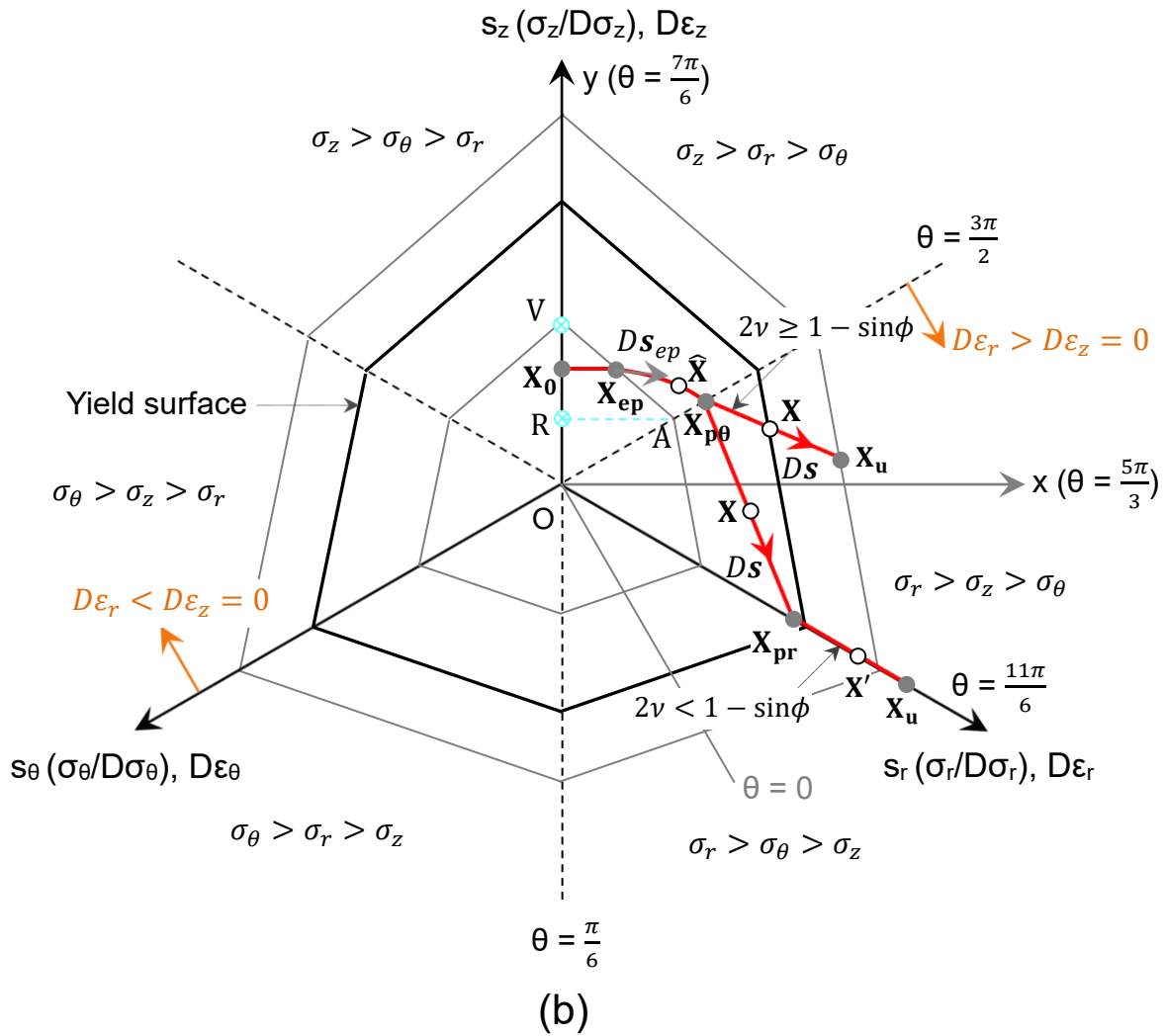

**Fig. 2. (Cont'd)** Graphical representation of stress state/paths in deviatoric plane for a soil element during cavity expansion process with $K_0 < 1$: (a) $K_{0(R)} < K_0 < 1$; (b) $K_{0(V)} < K_0 < K_{0(R)}$

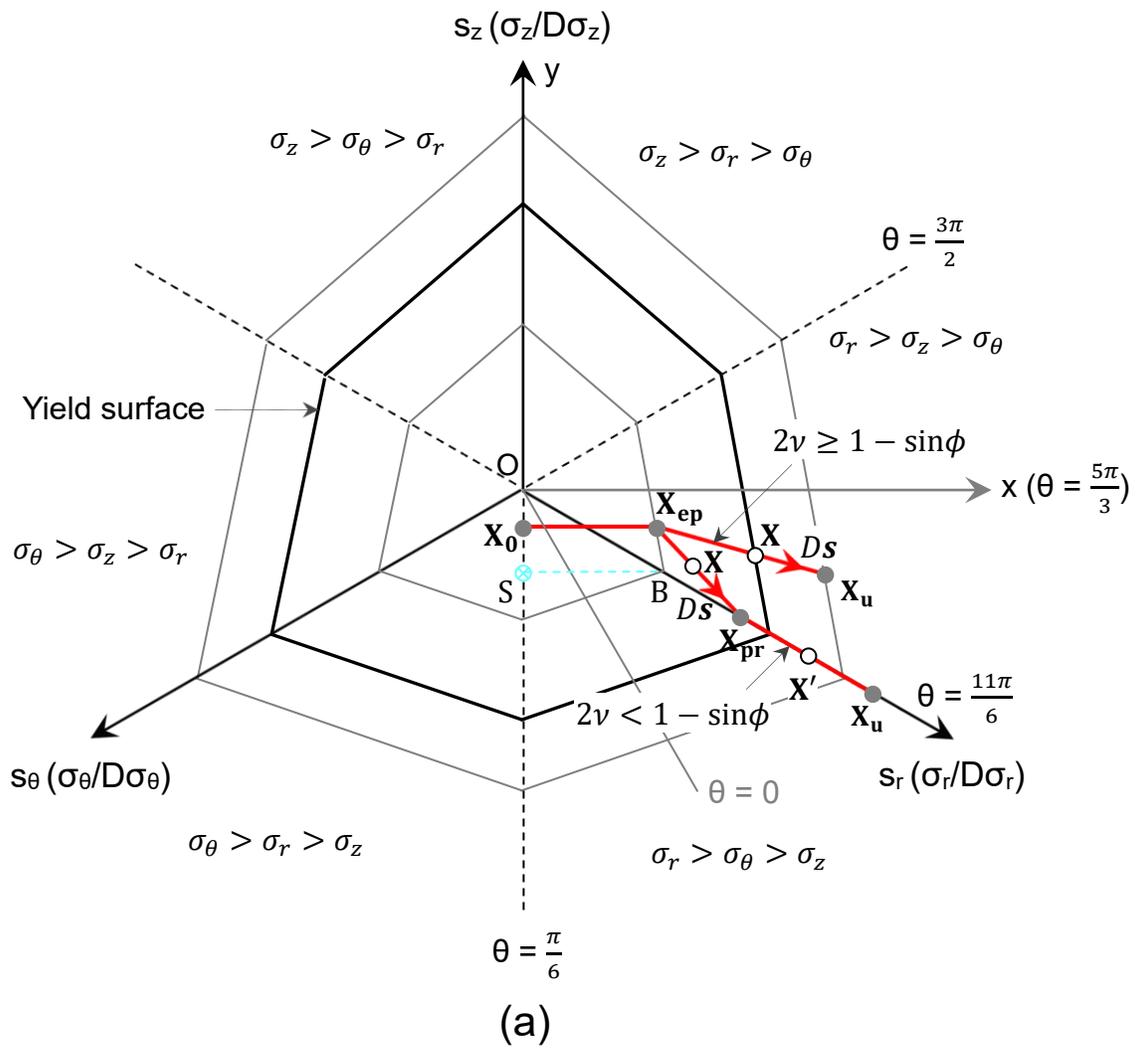

Fig. 3. Graphical representation of stress state/paths in deviatoric plane for a soil element during cavity expansion process with $K_0 > 1$: (a) $1 < K_0 < K_{0(S)}$; (b) $K_{0(S)} < K_0 < K_{0(T)}$

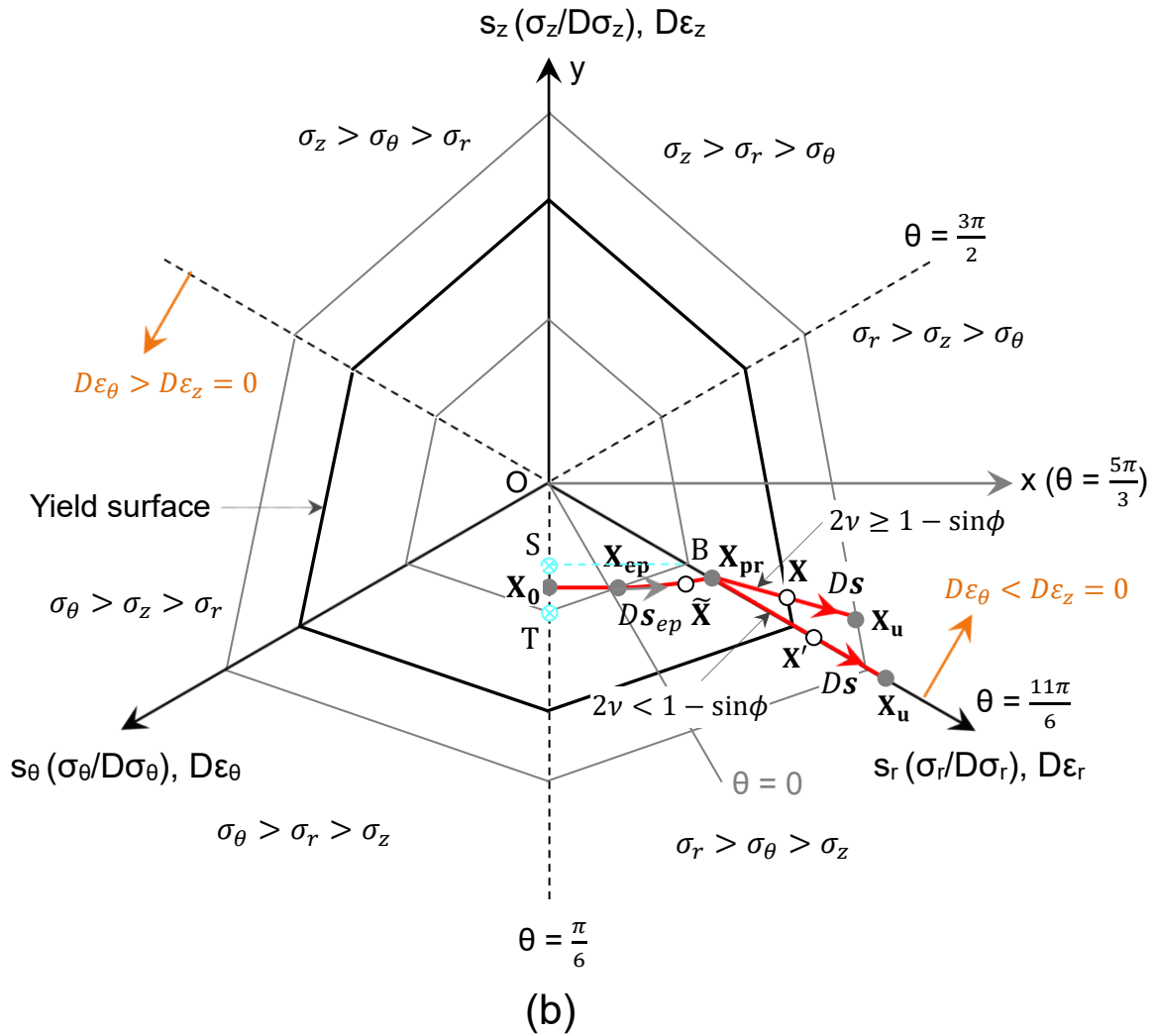

Fig. 3. (Cont'd) Graphical representation of stress state/paths in deviatoric plane for a soil element during cavity expansion process with $K_0 > 1$: (a) $1 < K_0 < K_{0(S)}$; (b) $K_{0(S)} < K_0 < K_{0(T)}$

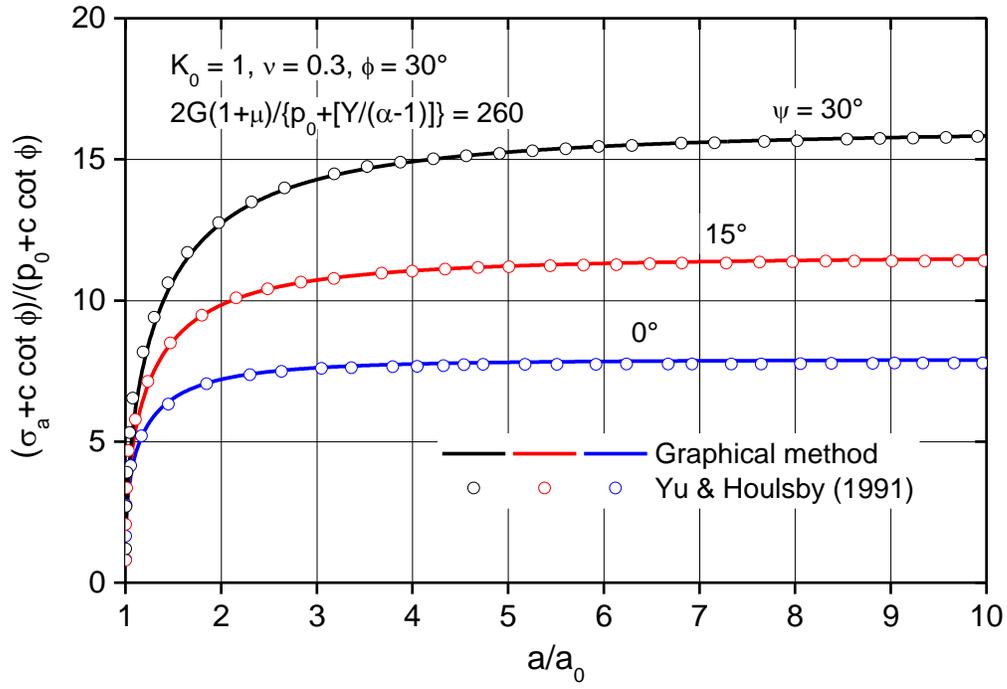

**Fig. 4. Comparisons of cavity expansion curves between current graphical method and Yu & Houlsby (1991) for the special case of $2\nu > 1 - \sin\phi$ with $K_0 = 1$**

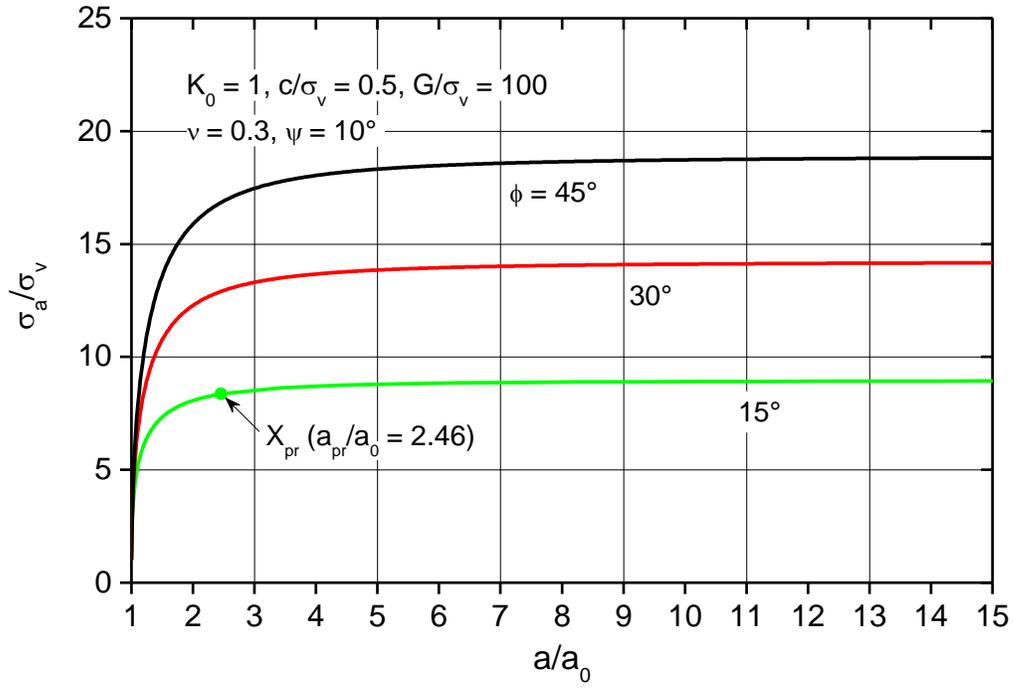

**Fig. 5.** Influences of friction angle on the variation of normalized internal cavity pressure with expansion ratio, $K_0 = 1$

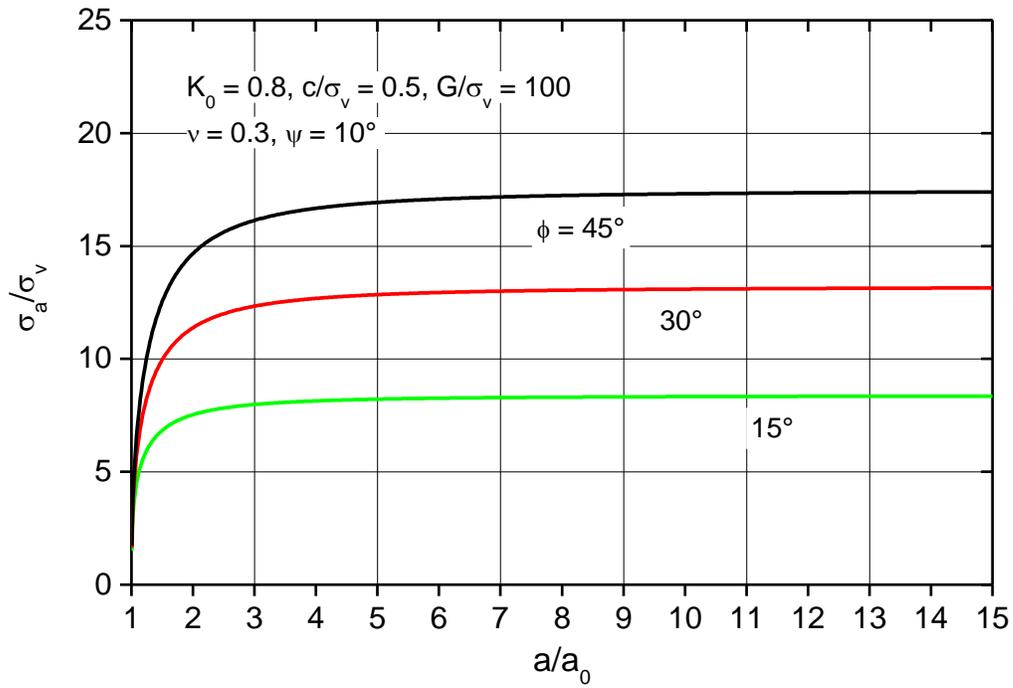

**Fig. 6.** Influences of friction angle on the variation of normalized internal cavity pressure with expansion ratio, $K_0 = 0.8$

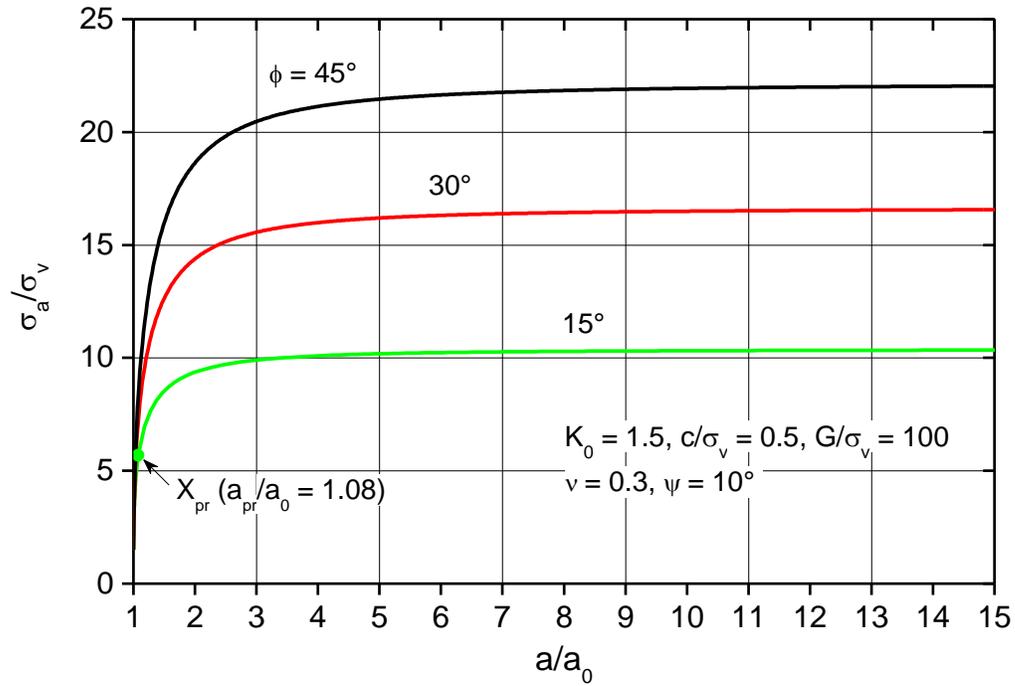

**Fig. 7.** Influences of friction angle on the variation of normalized internal cavity pressure with expansion ratio, $K_0 = 1.5$

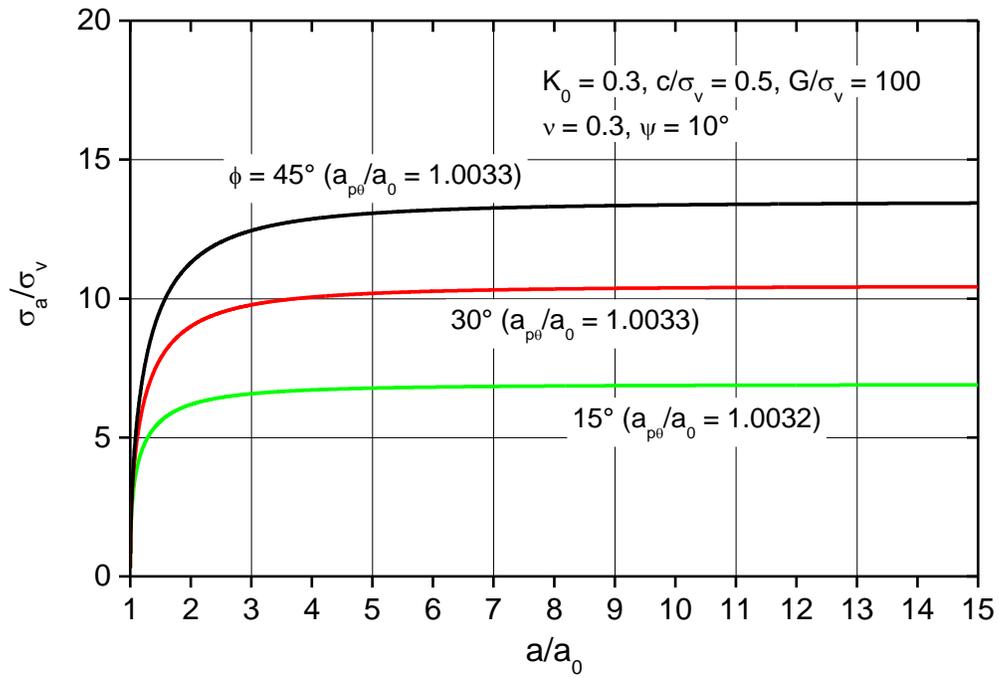

**Fig. 8. Influences of friction angle on the variation of normalized internal cavity pressure with expansion ratio, $K_0 = 0.3$**

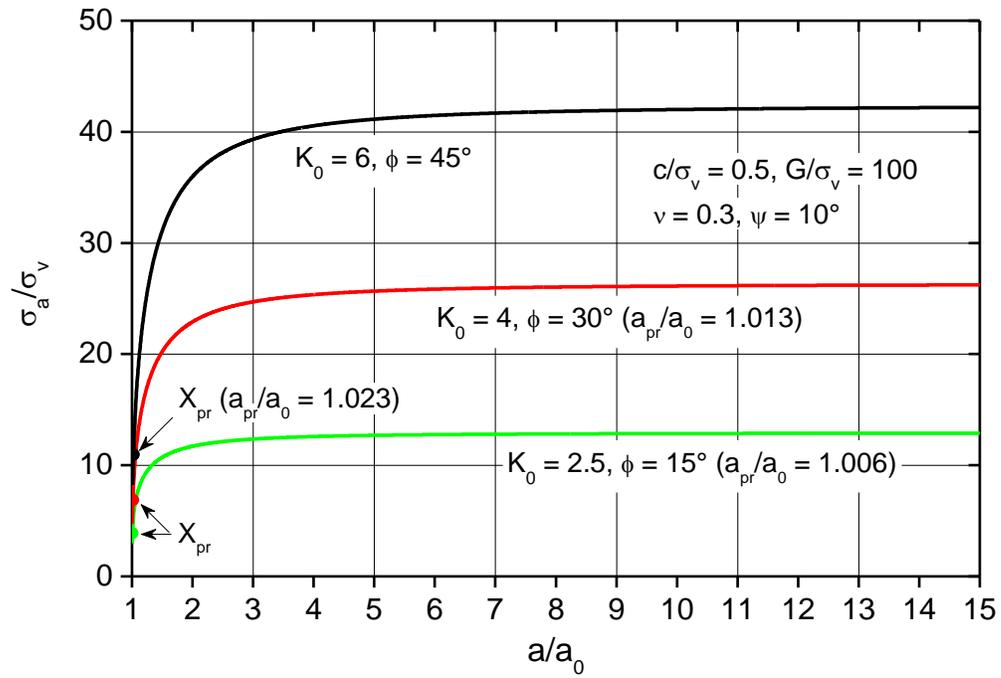

**Fig. 9. Influences of friction angle on the variation of normalized internal cavity pressure with expansion ratio, $K_0 = 2.5$; $K_0 = 4$; and $K_0 = 6$**

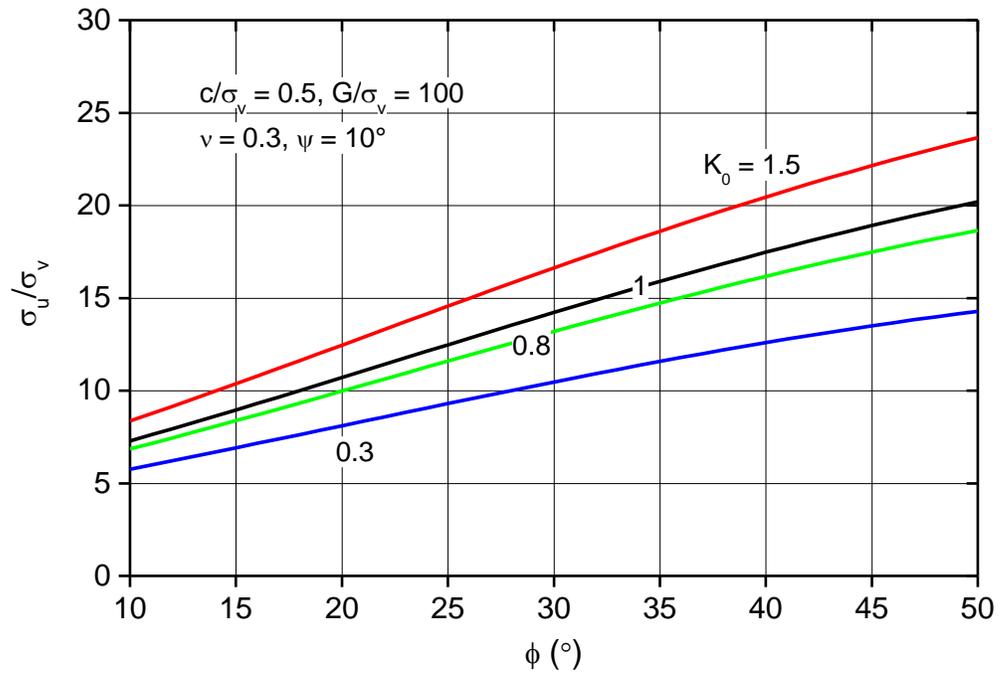

**Fig. 10. Limit cavity pressure in variation with friction angle for various $K_0$ values**